\documentclass[12pt, reqno]{amsart}
\usepackage{xr}
\usepackage{amsthm}
\usepackage[mathscr]{eucal}
\usepackage{amsfonts}
\usepackage{amsmath}
\usepackage{amssymb}
\usepackage{graphicx}
\usepackage{pstricks}
\usepackage{fullpage}
\usepackage{hyperref}
\usepackage{booktabs}
\usepackage{colortbl}
\usepackage{caption}
\usepackage{subcaption}
\usepackage[foot]{amsaddr}
\usepackage{setspace}
\usepackage{tabularx}
\usepackage{natbib}
\usepackage{multibib}
\usepackage{color}
\usepackage{tikz}

\usepackage{tabularx,ragged2e,booktabs,caption}
\newcolumntype{C}[1]{>{\Centering}m{#1}}

\newcommand\solidrule[1][0.5cm]{\rule[0.5ex]{#1}{1.8pt}}
\newcommand\dashedrule{\mbox{%
  \solidrule[2mm]\hspace{1mm}\solidrule[2mm]\hspace{1mm}\solidrule[2mm]}}

\newcommand\dotsrule{\mbox{%
  \solidrule[0.5mm]\hspace{1mm}\solidrule[0.5mm]\hspace{1mm}\solidrule[0.5mm]\hspace{1mm}\solidrule[0.5mm]}}

\newrgbcolor{purple}{0.64 0.21 0.61}  
  
\newcommand{\mb}[1]{\mathbb{#1}}
\newcommand{\mc}[1]{\mathcal{#1}}
\newcommand{\msc}[1]{\mathscr{#1}}
\newcommand{\bs}[1]{\boldsymbol{#1}}

\newcommand{\msf}[1]{\mathsf{#1}}

\newcommand{\tn}[1]{\textnormal{#1}}
\newcommand{\inner}[2]{\langle #1,#2\rangle}

\newcommand{\B}[1]{{\bf #1}}
\newcommand{\ind}{{1\hspace{-2.5pt}\tn{l}}}

\DeclareMathOperator*{\argmin}{argmin}

\newcommand{\R}{\mb{R}}

\newcommand{\N}{\mb{N}}

\newcommand{\prob}{\B{P}}

\newcommand{\fancyS}{\msc{S}}

\newcommand{\wh}[1]{\widehat{#1}}
\newcommand{\wt}[1]{\widetilde{#1}}

\newcommand{\vt}{\vartheta}
\newcommand{\ME}{\msf{E}}
\newcommand{\VAR}{\msf{VAR}}
\newcommand{\MSE}{\msf{MSE}}
\newcommand{\COV}{\msf{COV}}

\newcommand{\BIAS}{\msf{BIAS}}

\newcommand{\JUSD}{\operatorname{JUSD}}
\newcommand{\SMUCE}{\operatorname{SMUCE}}

\newenvironment{pf}[1][Proof] {\begin{proof}[\textit{\textbf{#1}}]} {\end{proof}}
\theoremstyle{plain}
\newtheorem{theo}{Theorem}
\newtheorem{lem}{Lemma} 
\newtheorem{prop}{Proposition} 
\newtheorem{cor}{Corollary}
\theoremstyle{definition}

\newtheorem{rmk}{Remark}
\newtheorem{alg}{Algorithm}
\numberwithin{equation}{section}
\title[Difference-based autocovariance estimation in regression]{Autocovariance 
estimation in regression with a discontinuous signal
and $m$-dependent errors: A difference-based approach}
\author{Inder Tecuapetla-G\'omez$^{(1)}$}
\email[Inder~Tecuapetla]{itecuap@mathematik.uni-goettingen.de}
\author{Axel Munk$^{(1,2)}$}
\address{$^1$Institute for Mathematical Stochastics\\
	University of G\"ottingen\\
	Goldschmidtstrasse 7, 37077 G\"ottingen}

\address{$^2$Max Planck Institute for Biophysical Chemistry\\
	Am Fassberg 11, 37077 G\"ottingen}
\email[Axel~Munk]{munk@math.uni-goettingen.de}
\date{\today}
\begin{document}
%
\begin{abstract}
 We discuss a class of difference-based estimators for the autocovariance in 
 nonparametric regression when the signal is discontinuous (change-point regression),
 possibly highly fluctuating, and the errors form a stationary $m$-dependent 
 process.
 These estimators circumvent the explicit pre-estimation of the unknown regression 
 function, a task which is particularly challenging for such signals.
 We provide explicit expressions for their mean squared errors
 when the signal function is piecewise constant (segment regression) and the 
 errors are Gaussian. Based on this we derive biased-optimized estimates
 which do not depend on the particular (unknown) autocovariance
 structure. Notably, for positively correlated errors, that part
 of the variance of our estimators which depends on the signal is minimal as well.
 Further, we provide sufficient conditions for $\sqrt{n}$-consistency;
 this result is extended to piecewise H\"older regression
 with non-Gaussian errors.
 
 We combine our biased-optimized autocovariance estimates with a projection-based
 approach and derive covariance matrix estimates, a method which is of 
 independent interest. Several simulation studies as well as an application
 to biophysical measurements complement this paper. 
\end{abstract}
\keywords{Autocovariance estimation, change-points, convex projection,
covariance matrix estimation, difference-based methods, discontinuous signal,
$m$-dependent processes, mean squared error, nonparametric regression}
%
%
\maketitle
%
%
%




\section{Introduction}~\label{sec:Introduction}

%

In nonparametric regression with correlated errors,
\begin{equation}\label{eq.NPR}
 y_i = f(x_i) + \varepsilon_i,\quad i=1,\ldots,n,
\end{equation}
where $(x_i)$ are the sampling points, $f$ is an unknown mean function or \emph{signal}, 
and $(\varepsilon_i)$ are zero mean stationary time series errors, 
the autocovariance $\gamma_h=\ME[\varepsilon_1\,\varepsilon_{1+h}]$,
$h=0,1,\ldots$, plays a prominent role for various tasks.
When the signal $f$ is smooth, the autocovariance appears, e.g., in the 
asymptotic variance of kernel estimators of $f$ and is important
for bandwidth selection and for inferential procedures, cf.~\cite{Opsomer.etal.01}.
In general, knowledge of the autocovariance, and in particular of the 
variance $\sigma^2 = \gamma_0$, is required for efficient signal estimation, e.g.~in 
wavelet-based estimation the autocovariance can be used for 
improved thresholding of the empirical wavelet coefficients, cf.~\cite{Johnstone.Silverman.97},
\cite{VonSachs2000}, \cite{Kovac2000}. 
Additionally, when the signal is discontinuous, which will be considered in this paper,
the autocovariance function is required to provide efficient estimates and confidence
regions for the location of the
discontinuities as well as the magnitude of their corresponding jumps, 
see Section~\ref{sec:CPapp} for an example.
Autocovariance estimation under a discontinuous and potentially highly fluctuating signal, 
however, is a notoriously difficult task in general and some knowledge about the dependence
structure is necessary. Therefore, throughout this paper we will consider 
zero mean, stationary, $m$-dependent errors, i.e., $\gamma_h = 0$ for $|h| > m$;
$\gamma_m \neq 0$. Although $m$-dependency ensures that the autocovariance function is 
zero starting at lag $m + 1$, this characteristic of our error model can often
be construed as a convenient proxy to more general situations, e.g.~when 
the autocovariance function decays exponentially with increasing lag.

\begin{figure}[htb]
\centering
  \scalebox{1}{
  \begin{subfigure}{0.45\linewidth}
    \centering
    \includegraphics[width=\linewidth]{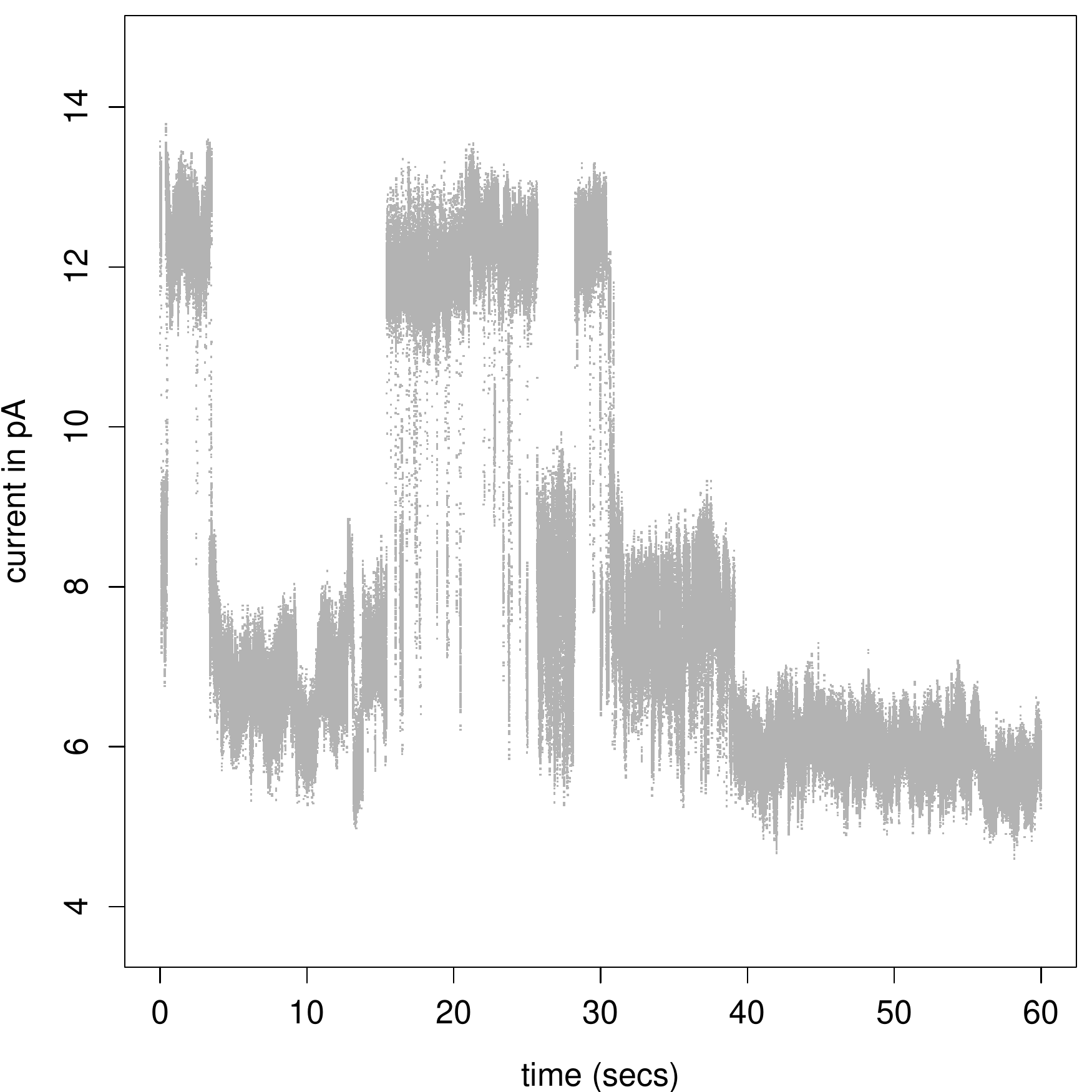} 
    \caption{} 
    \label{figIntro:a} 
  \end{subfigure} 
  \hfill
  \begin{subfigure}[H]{0.45\linewidth}
     \centering
     \includegraphics[width=\linewidth]{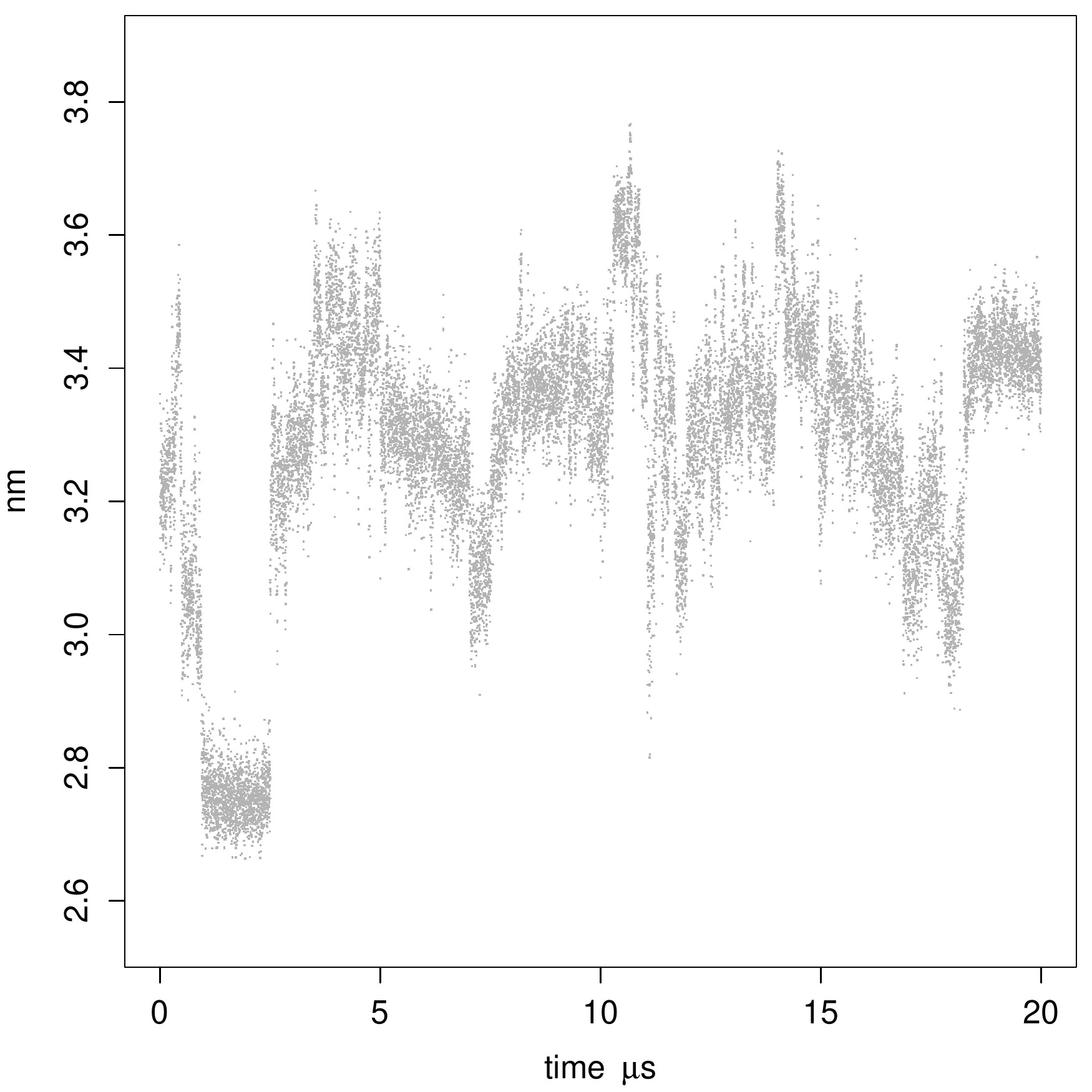} 
     \caption{} 
     \label{figIntro:b} 
   \end{subfigure}}
  \caption{\footnotesize{(A) 60 s of gramicidin A, see Section~\ref{sec.Applications} 
   for further details about this dataset.
   (B) 20000 ns of the trajectory 
   of the distance between a backbone atom and the first center of mass of 
   water channel AQY1, cf.~\cite{Krivobokova.etal.12} for further details.}}
  \label{fig:FigureIntro} 
\end{figure}

Regression models with discontinuous signal and $m$-dependent errors
as considered in this paper are of relevance in several areas of application. 
Figure~\ref{fig:FigureIntro} displays time series from two common biophysical measurements:
(A) recordings of an ion channel trace and (B) the 
trajectory of a molecular dynamic protein. 
For (A), the mean is typically modelled with a \emph{locally constant signal}
(also called \emph{change-point segment regression}) according to the openings and closings 
of the channel, cf.~\cite{VanDongen.96}, and $m$-dependence results from the 
low-pass filter utilized to digitize ion channel measurements, cf.~\cite{Hotz.etal.13}.
For (B) a more flexible signal assumption (\emph{piecewise smooth change-point regression})
seems in order and often $m$-dependence can be confirmed empirically.

The contributions of this paper are also relevant for change-point estimation
and detection which have been investigated extensively in the particular case of independent 
errors, see e.g.~\cite{Page.54, Page.55}, \cite{Dumbgen.91}, \cite{Brodsky.Darkhovsky.93} and
\cite{Carlstein.94} for some early references.
More recently, and arguably motivated by large scale applications, e.g.~from genetics,
the focus has been on the recovery of signals with a potentially large number of 
change-points, see e.g.~\cite{Olshen.etal.04}, \cite{Fearnhead.Liu.07}, \cite{Spokoiny.09},
\cite{Harchaoui.LevyLeduc.10},
\cite{Killick.etal.12}, \cite{Siegmund.13}, 
\cite{Frick.Munk.Sieling.14}, \cite{Du.15} and \cite{Li.etal.16}
among many others. For serially correlated errors, 
change-points estimates have been primarily investigated 
asymptotically and when the number of change-points
is finite (but unknown) see e.g.~\cite{Davis.etal.06},
\cite{Fryzlewicz.14}, \cite{Preuss.etal.14} and \cite{Chakar.etal.16}. 
To some extent, such an asymptotic analysis resembles finite sample situations
when the number of change-points is small (relative to the
sample size) and variance-covariance estimation
becomes less cumbersome (\cite{Picard.85}, \cite{Huvskova.07}).
Autocovariance estimation may even be disregarded for consistent change-point estimation
over large classes of dependency structures (\cite{Lavielle2000}, \cite{Bardet2012}).

In contrast, in this paper we mainly adopt a non-asymptotic perspective 
motivated by highly fluctuating signals having a potentially large number 
of change-points as it is known that this will increase the finite sample bias 
of any standard variance-covariance estimate.
Hence, subsequent use of these biased estimates may significantly 
reduce the efficiency of change-point statistics, cf.~Eqs.~(13)-(15) 
of \cite{Jandhyala.13}.
As we will show later on, even when the number of change-points increases
with the sample size, it is still possible to account for bias-reducing and 
$\sqrt{n}$-consistent estimates for $\gamma_{(\cdot)}$.
This is reflected by the fact that estimation of $\gamma_{(\cdot)}$ is simpler than 
that of the entire signal $f$, which is well-known for $\gamma_0 = \sigma^2$ in 
the independent case (e.g.~\cite{Spokoiny.2002}).

In summary, in those situations when the signal fluctuation becomes
dominant (which will be made precise later on), pre-estimation
of the signal $f$ is notoriously difficult and direct estimation
of $\gamma_{(\cdot)}$ becomes pertinent.
This paradigm is well known, in particular, for independent noise, 
i.e.~for estimation of the variance $\sigma^2$. 
For this task, \emph{difference-based estimators} provide a simple and 
practical solution:
a difference sequence $\{\Delta_i\}$ is a sequence of real numbers such that 
\begin{equation}\label{eq.Deltas}
 \sum\,\Delta_i = 0, \quad \sum \Delta_i^2 = 1.
\end{equation}
Assume that $\Delta_i = 0$ for $i < -l_1$ and $i > l_2$, and $\Delta_{-l_1}\,\Delta_{l_2}\neq 0$
with $l_1, l_2 \geq 0$. Then $l = l_1 + l_2$ is called the \emph{order} of the
sequence; usually $l_1 = 0$ and $l_2 = l$. 
Following \cite{Hall.etal.90} a difference-based estimate
of $\sigma^2$ has the form
\begin{equation}\label{eq.sigma-hat}
 \hat{\sigma}^2 = (n-l)^{-1}\,\sum_{k = l_1 + 1}^{n - l_2}\left( \sum_i\, \Delta_i\, y_{i + k} \right)^2. 
\end{equation}
When the signal is smooth, estimators of this type have been investigated extensively
see e.g.~\cite{Rice.84}, \cite{Gasser.etal.86}, \cite{Muller.Stadtmuller.87},
\cite{Dette.etal.98}, \cite{Spokoiny.2002}, \cite{Brown.Levine.2007}, 
\cite{Tong.etal.13}, \cite{Dai.etal.2015}.
As argued by \cite{Munk.05} and others, a particular appeal of these estimators
is that their weights can be adapted to high fluctuation 
of the signal, i.e.~for bias reduction.
As we will see in this paper, this feature makes difference-based
estimators particularly useful also in applications with correlated errors
where the signal exhibits high fluctuation and discontinuities.

Difference-based estimators have also been used in
nonparametric regression with stationary errors.
For instance, \cite{Muller.Stadtmuller.88} proposed estimators based on differences 
of first order to estimate (invertible) linear transformations of the variance-covariance 
matrix of stationary $m$-dependent errors. \cite{Herrmann.92} suggested differences 
of second order to estimate the zero frequency of the spectral density 
of stationary processes with short-range dependence.
For autoregressive errors, \cite{Hall.VanKeilegom.03} proposed $\sqrt{n}$-consistent 
and, under normality, efficient autocovariance estimates.
Under some mixing conditions, \cite{Park.etal.06} suggested 
to estimate the autocovariance function applying difference-based estimators of first order 
to the residuals of a kernel-based fit of the signal.
Most close to our work is \cite{Zhou.etal.15}, who provide
an optimal difference-based estimate of the variance $\sigma^2 = \gamma_0$ 
for smooth nonparametric regression when the errors are correlated.
Their optimized weights, however, depend on the remaining values of the 
autocovariance function, i.e.~$\gamma_{h}$, $h \neq 0$, which in general
are unknown. In contrast, in this paper we estimate the entire autocovariance function
and our estimates depend solely on $m$. 
All the methods just discussed have been analyzed
for smooth signals and to the best of our knowledge, derivation of
optimal weights for autocovariance difference-based estimates
in the case of a discontinuous and highly fluctuating signal still remains elusive
and becomes the main focus of our work.

Summarizing this paper, we begin by suggesting the use of difference-based estimators of the
autocovariance of the error process in nonparametric regression \eqref{eq.NPR}.
Then we show that for $m$-dependent errors we need to use differences of gap
$m+1$. We obtain finite-sample results for the mean squared error ($\MSE$) of these estimators.
This $\MSE$ includes a bias term which depends on the unknown regression
function $f$, in situations where $f$ fluctuates substantially this term could
dominate. We show that to minimize this bias term (and that part of the variance
which depends on $f$ as well) we should use estimators
based on only first or second order differences and give explicit forms
for the optimal choice of weights in the difference-based estimators.
Further, we provide sufficient conditions for $\sqrt{n}$-consistency;
this result is extended to piecewise H\"older regression with non-Gaussian 
errors. A more detailed account of these and other results is presented in 
the next section. The theory is complemented by simulation studies, a data example,
and statistical software for autocovariance estimation.

\section{Main results}~\label{sec:Results}

As a prototypical example of a regression model with a discontinuous signal
we consider for the moment \eqref{eq.NPR} with a signal $f$ that is locally 
constant (\emph{change-point segment regression}) and hence admits the representation
\begin{equation}\label{eq.piecewise.function}
 f(x) = \sum_{j=0}^{K-1}\,a_j\ind_{[\tau_j,\tau_{j+1})}(x),\quad x\in[0,1),\quad a_j\neq a_{j+1}.
\end{equation}
Here the change-points of $f$, $0=\tau_0<\tau_1<\cdots<\tau_{K-1}<\tau_K=1$, 
its levels $(a_j)_{0\leq j\leq K-1}$, and the number of discontinuities $K\in \N$ 
are unknown and can be potentially large.
Since for large enough $K$ any discretized function can be represented as in 
\eqref{eq.piecewise.function}, this equation describes a wide class of signals.
For simplicity of presentation we will assume that
the sampling points $(x_i)$ are equally spaced on $[0,1)$, i.e., $x_i=i/n$. 
Further, $f_i$ will denote $f(x_i)$.
 
\subsection{Autocovariance estimation}\label{sec:Intro.Pointwise}

 Now we introduce the class of difference-based estimators to be considered
 for $m$-dependent errors. 
 Let ${\bf Y}$ denote the vector of observations $y_i$ following \eqref{eq.NPR}.
 For $1\leq lh< n$, a \emph{generalized difference-based estimator of order $l$
 and gap $h$} is a random quadratic form
 \begin{equation}\label{eq.DBEk}
  {\bf Q}_h({\bf Y},\bs{w}_l) 
  = 
  \frac{1}{P(\bs{w}_l)(n-lh)}
  \sum_{i=1}^{n-lh}(d_0y_i + d_1y_{i+h}+ d_2y_{i+2\,h} \ldots+d_l\,y_{i+l\,h})^2,
 \end{equation}
 where $\bs{w}_l = ( d_0 \quad d_1 \quad \cdots \quad d_l )^\top\in\R^{l+1}$
 is a vector of numbers (weights) satisfying
 \begin{equation}\label{eq.a}
    \sum_{i=0}^{l}\,d_i=0,
 \end{equation}
 and $P(\bs{w}_l) = \sum_{i=0}^h d_i^2$.
  Setting $h=1$ in \eqref{eq.DBEk} we get difference-based
  estimators of order $l$, cf.~Eq.\eqref{eq.sigma-hat}. 
 For $l=1$ and $\bs{w}_{1} = (1 \quad -1)^\top\in\R^{2}$,
 $P(\bs{w}_{1})=2$ and for $1\leq h < n$ we get the ordinary difference-based estimator 
 of gap $h$ (for $h=1$ see \cite{Rice.84}):
 \begin{equation}\label{eq.odb.h}
  \wh{\delta}^{(h)}
  =
  \frac{1}{2(n-h)}\,
  \sum_{i=1}^{n-h}\left( y_i - y_{i+h}  \right)^2.
  \end{equation} 
  
  Note that for independent errors in \eqref{eq.NPR}, 
  $\ME[\wh{\delta}^{(h)}] = \gamma_0 + o(1)$
  provided $\sum_{i=1}^{n-h}(f_i-f_{i+h})^2 = o(n)$ and we can 
  use the estimator \eqref{eq.odb.h} to get asymptotically unbiased
  estimates of the variance. In contrast, for stationary errors the
  situation becomes more complicated as then
  $\ME[\wh{\delta}^{(h)}] = \gamma_0-\gamma_h + o(1)$.

  We will show for the change-point segment model~\eqref{eq.piecewise.function}
  (Theorem~\ref{theo.NoGoal} of Section~\ref{sec.NoGoalTheo})
  that for $m$-dependent errors any estimator of the variance $\gamma_0 = \sigma^2$
  whose $\MSE$ tends to 0
  and that satisfies \eqref{eq.DBEk}-\eqref{eq.a} must have 
  gap $h$ at least $m+1$.
  Therefore, we will focus on this type of variance estimators in the following. 
  We will further restrict ourselves to estimators of first or second order,
  see Lemma~\ref{lem.lim.DF.PC} below for justification of this. 
 We commence now a discussion about the $\MSE$ of this class of estimators.
  
 Let $n_m=n-2(m+1)$ and w.l.o.g.~set $d_0=1$, then estimators for $\gamma_0$
 based on differences of second order ($l=2$) and gap $m+1$ in 
 \eqref{eq.DBEk} can be written as
 \begin{equation}~\label{eq.gamma0}
   \wh{\gamma}^{(m)}_0(d) 
    = 
   \frac{(1+d+d^2)^{-1}}{2\,n_m}\,
    \sum_{i=1}^{n_m}\left( y_i -(1+d)y_{i+(m+1)} + d\,y_{i+2(m+1)} \right)^2,\quad d\in \R,
 \end{equation}
 using \eqref{eq.a}.
  Combining the ordinary difference-based estimator of gap $h$, cf.~\eqref{eq.odb.h}, 
  with \eqref{eq.gamma0} we will estimate the remaining values of the autocovariance function
  $\gamma_h$, $h=1,\ldots,m$. Namely, 
 \begin{equation}~\label{eq.gammak}
  \wh{\gamma}_h^{(m)}(d) 
  = 
  \wh{\gamma}^{(m)}_0(d)  - \wh{\delta}^{(h)},
  \quad d\in \R,
  \quad h = 1,\ldots,m.
 \end{equation} 
 Here $d\in \R$ is a parameter to be optimized later on.
 In order to present our first result we need to introduce the quadratic
 variation of $f$ in \eqref{eq.piecewise.function}:
 \begin{equation}~\label{eq.QuadraticVariation}
  J_K
  :=
  \sum_{j=0}^{K-1}(a_{j+1}-a_j)^2.
 \end{equation}
 
\begin{theo}\label{theo.MSE}
 Suppose that in the \emph{segment} regression model~\eqref{eq.NPR}-\eqref{eq.piecewise.function}
 the noise $(\varepsilon_i)_{1\leq i \leq n}$ is a sample from a zero mean, $m$-dependent, 
 stationary Gaussian process with autocovariance function $\gamma_h=\ME[\varepsilon_1\,\varepsilon_{1+h}]$, 
 $h=0,\ldots,m$. Additionally, assume that the change-points of $f$, 
 $0=\tau_0<\tau_1<\cdots<\tau_{K-1}<\tau_K=1$, satisfy that
 \begin{equation}\label{eq.DistBetweenJumps}
  \min_{1\leq i\leq {K-1}}| \tau_{i+1} - \tau_{i} | > 4(m+1)/n.
 \end{equation}

 Then, for $m\geq 1$
 \begin{equation}~\label{eq.MSE1}
 \MSE[ \wh{\gamma}_0^{(m)}(d) ]
  =
  \underbrace{n^{-2}\,p_0^2(d)\,J_K^2}_{\BIAS^2}
  + 
  \underbrace{n^{-2}\left( p_1(d;\gamma_{(\cdot)})\,\gamma_0\,J_K + n\,p_2(d;\gamma_{(\cdot)}) + 
	p_3(d;\gamma_{(\cdot)}) \right)}_{\VAR}, 
 \end{equation}
  and for $h=1,\ldots,m$
  \begin{equation}~\label{eq.MSE2}
    \MSE[ \wh{\gamma}_h^{(m)}(d) ]
  =
  \underbrace{n^{-2}\,(p_0^\ast)^2(d)\,J_K^2}_{\BIAS^2}
  +
  \underbrace{n^{-2}\left( p_1^\ast(d;\gamma_{(\cdot)})\,J_K + n\,p_2^\ast(d;\gamma_{(\cdot)}) + 
  p_3^\ast(d;\gamma_{(\cdot)}) \right)}_{\VAR}.
  \end{equation}
\end{theo}  
 See Section~\ref{sec.bias.minimizer} for explicit expressions
 of $p_0$, $p_0^\ast$, $p_1$, $p_1^\ast$, $p_2$, $p_2^\ast$, $p_3$
 and $p_3^\ast$. We remark that Gaussianity is only needed to get 
 the variance explicitly. 
 
 \begin{rmk}
 As mentioned above for simplicity we are considering equally 
 spaced observations sampled from the unit interval at rate $1/n$.  
 Note, however, that our results can be transferred to general sampling
 points as we provide the finite sample $\MSE$s \eqref{eq.MSE1}-\eqref{eq.MSE2}
 and the only restriction is \eqref{eq.DistBetweenJumps} which
 transfers to general sampling points by replacing $\tau_j$ by $\lfloor n \tau_j\rfloor$,
 where $\lfloor x \rfloor$ denotes the integer part of $x$.
 \end{rmk}

 Now let us discuss Theorem~\ref{theo.MSE}. As an immediate
 consequence we deduce that the influence of $f$ over the $\MSE$s
 \eqref{eq.MSE1}-\eqref{eq.MSE2} only appears through its quadratic variation 
 $J_K$ \eqref{eq.QuadraticVariation}.
 Let us first consider \eqref{eq.MSE1}. 
 Here the bias consists of $n^{-1}\,p_0(\cdot)$, a rational function, times $J_K$.
 We can further define the \emph{extended bias}, $\BIAS^\ast$, of $\wh{\gamma}_0^{(m)}(d)$
 as the part of the $\MSE$ in \eqref{eq.MSE1} that depends on $f$:
 \begin{equation}\label{eq.ExtBias}
  \BIAS^\ast[\wh{\gamma}_0^{(m)}(d)] := n^{-2} \left[p_0^2(d)\,J_K^2 + p_1(d;\gamma_{(\cdot)})\,\gamma_0\,J_K\right].
 \end{equation}
 In order to justify our definition of extended bias, note that
 unlike $p_1$, the rational functions $p_2(\cdot;\gamma_{(\cdot)})$
 and $p_3(\cdot;\gamma_{(\cdot)})$ do not align with $f$
 and these functions affect the part of the variance which solely
 depends on the autocovariance. 
 A similar analysis and conclusions can 
 be made for the estimators $\wh{\gamma}_h^{(m)}(d)$ in \eqref{eq.MSE2}.

 In view of \eqref{eq.MSE1} and \eqref{eq.MSE2} we deduce that
 for highly oscillating and discontinuous signals, in \eqref{eq.piecewise.function},
 e.g.~for large $K$, the influence of $J_K$ over the $\MSE$s may become dominant and 
 might not be negligible even for very large sample sizes.
 On the contrary, the influence of the unknown autocovariance function $\gamma_{(\cdot)}$
 over the $\MSE$s is comparably small in this situation.

  In light of the above we focus on finding variance estimates in \eqref{eq.gamma0}
 and autocovariance estimates in \eqref{eq.gammak} whose $\MSE$s (see Theorem~\ref{theo.MSE}) 
 have minimal influence from $J_K$ (and hence from $f$). Our main results on autocovariance
 estimation for change-point regression with stationary $m$-dependent errors
 are stated now.\smallskip
  
 
 {\bf Estimation of $\gamma_0$.}
 From Theorem~\ref{theo.MSE} we get that
 $\BIAS[ \wh{\gamma}_0^{(m)}(d) ]=n^{-1}p_0(d)\,J_K$, see \eqref{eq.MSE1}. 
 Since for $m\in\N$, $\min_{d\in\R}p_0(d)=p_0(1)=((m+1)/3)^2$, 
 for details see \eqref{eq.ExpValGamma0}, we have proven the following:
 \begin{theo}\label{cor.BIASgamma0}
  Let $\wh{\gamma}_0^{(m)}(d)$ be the difference-based estimator 
  given by \eqref{eq.gamma0}.
 Suppose that in the segment regression model~\eqref{eq.NPR}-\eqref{eq.piecewise.function}-\eqref{eq.DistBetweenJumps},
 the noise $(\varepsilon_i)_{1\leq i \leq n}$ is a sample from a zero mean, $m$-dependent, 
 stationary process. Then for $m\geq 1$,
  $\BIAS[\wh{\gamma}_0^{(m)}(d)]$ is minimized at $d=1$, i.e., by the 
  estimator
 \begin{equation}
  \wh{\gamma}_0^{(m)}(1)
  =
  \frac{1}{6\,n_m}\,\sum_{i=1}^{n_m}\left( y_i - 2\,y_{i+(m+1)} + y_{i+2(m+1)} \right)^2,
  \quad
  n_m=n-2(m+1).\label{eq.Delta(121).1}
 \end{equation}
  \end{theo}
  Note that in Theorem~\ref{cor.BIASgamma0} a normal error assumption
  is not needed.
  Already \cite{Herrmann.92} have suggested this estimator in the context
  of nonparametric regression with a smooth signal and stationary errors.  
  Note that Theorem~\ref{cor.BIASgamma0} provides further justification
  for its use particularly when the signal is discontinuous.
  In fact, assuming further that the correlation is non-negative,
  we found that even the \emph{extended bias} of $\wh{\gamma}_0^{(m)}(d)$,
  is minimized at $d=1$. More precisely we get the following
  result whose proof can be found in Section~\ref{sec.bias.minimizer}.
 \begin{theo}\label{theo.Bias.diff-based.MAq}
 Suppose that the conditions of Theorem~\ref{theo.MSE} hold.
 Assume, additionally, that the autocovariance function of the noise
 $(\varepsilon_i)_{1\leq i \leq n}$ belongs to either of the
 following classes:
 \begin{enumerate}
   \item $\gamma_h=\rho\,\gamma_0$, for $h=1,\ldots,m$ such that 
    \begin{equation}\label{eq.RegionMaximalCorrelation}
    1+2\,\rho\,\sum_{h=1}^{m}\,\cos(h\,\lambda)\geq 0,\quad\forall \lambda\in [-\pi, \pi].
    \end{equation}
     
   \item $\gamma_h\geq 0$ for $h=1,\ldots,m$.      
  \end{enumerate}
  Then for $m\geq 1$, the estimator $\wh{\gamma}_0^{(m)}(1)$ in \eqref{eq.Delta(121).1}
  minimizes the extended bias $\BIAS^\ast[ \wh{\gamma}_0^{(m)}(d) ]$, cf.~\eqref{eq.ExtBias}.
 \end{theo}
 
 Eq.~\eqref{eq.RegionMaximalCorrelation} ensures that the function
 $\gamma_h = \rho\,\gamma_0$ for $h = 1,\ldots, m$ and zero otherwise, is the autocovariance of
 a zero mean, stationary, $m$-dependent process, cf.~Corollary~4.3.2 of \cite{Brockwell.Davis.06}.\smallskip
  
 {\bf Estimation of $\gamma_h$, $h = 1,\ldots,m$.}
 We will show that the following weights optimize the bias of estimators given by \eqref{eq.gammak}:
 \begin{equation}\label{eq.Delta(121).2}
  d_{h,m}
  =
 \begin{cases}
  1 & \mbox{ for } h < \frac{2}{3}(m+1)\\
  \frac{h \pm \sqrt{ h^2 - 4(m+1-h)^2 }}{2(m+1-h)} & \mbox{ otherwise } 
 \end{cases}.
 \end{equation}
  \begin{theo}\label{prop.Opt-W1}
  Suppose that the conditions of Theorem~\ref{cor.BIASgamma0} hold.  
 Then for $m\geq 1$, $\BIAS[\wh{\gamma}_h^{(m)}(d)]$ is minimized at $d_{h,m}$, 
  cf.~\eqref{eq.Delta(121).2}. In particular, for those values of $h$ such that 
  $h\geq \frac{2}{3}(m+1)$, $\wh{\gamma}_h^{(m)}(d_{h,m})$ is an unbiased estimate of $\gamma_h$.
 \end{theo}  
 \begin{pf}
    Since $\BIAS[\wh{\gamma}_h^{(m)}(d)]={p_0^\ast(d)}J_K/n$, cf.~Theorem~\ref{theo.MSE},
    where $p_0^\ast(d) = p_0(d)-h/2$, cf.~\eqref{eq.Qast0}, we only need 
    to minimize $p_0^\ast(\cdot)$ w.r.t.~$d$.
    For $3h \geq 2(m+1)$, $d_{h,m}$, cf.~Eq.~\eqref{eq.Delta(121).2}, is a root
    of $p_0^\ast$. For $3h<2(m+1)$, straightforward calculations yield
    that $d=1$ is a global minimum of $p_0^\ast$ on $\R$. This completes the 
    proof.
 \end{pf}

 Observe that the underlying autocovariance $\gamma_{(\cdot)}$ does not
 appear in the expression for $\wh{\gamma}_h^{(m)}(d_{h,m})$, $h=1,\ldots,m$.
 In contrast, the corresponding extended bias of $\wh{\gamma}_h^{(m)}(d)$, 
 $d\in \R$, depends on the unknown $\gamma_{(\cdot)}$ in an intricate 
 fashion, cf.~Eq.~\eqref{eq.MSE2}-\eqref{eq.Qast1}-\eqref{eq.Qast2}-\eqref{eq.Qast3}, 
 hence full minimization of this function is practically infeasible.
 
 In summary, we find that the difference-based estimates given by \eqref{eq.Delta(121).1} 
 and \eqref{eq.Delta(121).2} are bias-reducing for the autocovariance of 
 wide classes of stationary $m$-dependent processes and discontinuous signals. 
 Additionally, under Gaussianity and for non-negative correlation, \eqref{eq.Delta(121).1} 
 is even extended-bias-optimal. Moreover,
 Theorem~\ref{theo.AsympProperties.Gammah} of Section~\ref{sec:HolderRegression}
 establishes that for piecewise H\"older continuous signal (which includes segment
 regression \eqref{eq.piecewise.function}) 
 with general (non-Gaussian) errors whose all moments up to order 4 are stationary, 
 the estimates \eqref{eq.Delta(121).1}-\eqref{eq.Delta(121).2} are $\sqrt{n}$-consistent
 for $\gamma_{(\cdot)}$ as long as the number of discontinuities $K_n = o (\sqrt{n})$.
 
\subsection{Covariance matrix estimation}\label{sec:Intro.Covariance}

  Let $\wh{\Gamma}$ be the $n\times n$ symmetric Toeplitz matrix whose first
  $m + 1$ entries of its first row are filled with $\wh{\gamma}_{h}^{(m)}(d_{h,m})$, 
  $h=0,1,\ldots,m$, cf.~Eqs.~\eqref{eq.Delta(121).1}-\eqref{eq.Delta(121).2}, and
  the remaining $n - (m+1)$ entries are zero.
  As in general, \emph{pointwise} autocovariance estimates $\wh{\gamma}_h$, $h=0,1,\ldots,m$,
  may lead to a covariance matrix estimate which is not positive definite,
  cf.~\cite{Hall.VanKeilegom.03}, $\wh{\Gamma}$ may not be the exception.
  In order to overcome this problem we propose a projection-based estimator. 
  Define
 \begin{equation}\label{eq.CovMatEst}
  \wh{\Gamma}^\ast 
  = 
  P_{\mc{C}_n^{(m)}}(\wh{\Gamma})
  :=
  \argmin\{ \|\wh{\Gamma} - \Gamma^{(m)}\|_F: \Gamma^{(m)}\in\mc{C}_n^{(m)} \},
 \end{equation}
 that is, the unique projection of $\wh{\Gamma}$ onto $\mc{C}_n^{(m)}$, the closed
 convex set of all $n\times n$ symmetric, positive semidefinite, $(m+1)$-banded 
 Toeplitz matrices.
 In \eqref{eq.CovMatEst}, $\|\cdot\|_F$ denotes the Frobenius norm.
 We show that the \emph{projection-based}
 estimate $\wh{\Gamma}^\ast$ always has smaller mean squared error than the 
 \emph{pointwise-based} estimate $\wh{\Gamma}$, cf.~Theorem~\ref{theo.CovarianceEstimate}
 in Section~\ref{sec:CovMatrix}. This theorem
 might be of interest on its own since it relies on a general projection principle
 which can be applied to any ensemble of individual autocovariance estimators
 to obtain a symmetric, positive semidefinite Toeplitz (covariance) matrix estimate.
 Our final autocovariance matrix estimate $\wh{\Gamma}^\ast$ in \eqref{eq.CovMatEst}
 can be computed numerically by a Dykstra-type alternating projection algorithm
 which we will detail in Section~\ref{sec:AlternatingAlgorithm}.

\subsection{Numerical studies}

  Finite sample properties of our bias-reducing autocovariance estimators \eqref{eq.Delta(121).1}-\eqref{eq.Delta(121).2}
  are investigated in a series of simulations in Section~\ref{sec.Simulations} 
  and compared to other estimators from the literature:
  Section~\ref{sec:ParkErrors_Chakarsignal} assesses the performance of these
  estimates for autocorrelation estimation, their robustness against normal distributed 
  errors is also studied; robustness against the assumption of a piecewise constant signal is
  explored in Section~\ref{sec:ParkSetup}.
  When the signal is discontinuous we found that our estimates outperform all the others
  under comparison. For smooth signals and positive correlation our estimates 
  are competitive to some optimized kernel-based estimates.
   
\subsection{Applications}

In Section~\ref{sec:IonChannel} we analyze the dependence structure
of the data example shown in Figure~\ref{figIntro:a}. We found 
a $6$-dependent process as appropriate and estimate its autocorrelation. 
We found that is in close agreement to that obtained by theoretical
considerations from low-pass filtering.
Although in this paper we only considered an application on biophysical measurements, 
we stress that our method may proven useful for many other areas such as
the analysis of financial time series or sequential data in genetics.
Finally, in Section~\ref{sec:CPapp} we show how our estimates can be used
to improve considerably the estimation accuracy in the $m$-dependent case
of a change-point estimator initially designed for independent data.
 
\subsection{Software and Supporting Information}

The methods discussed in this paper are available in the \texttt{R} package \texttt{dbacf} 
(\href{http://www.stochastik.math.uni-goettingen.de/dbacf}{http://www.stochastik.math.uni-goettingen.de/dbacf}).
In this software we have implemented the estimates \eqref{eq.Delta(121).1}-\eqref{eq.Delta(121).2} 
(see function \texttt{dbacf})
as well as the alternating projection algorithm from Section~\ref{sec:CovMatrix} 
leading to $\wh{\Gamma}^\ast$ (see \texttt{nearPDToeplitz}).
We defer the proofs of most of our results to the Supporting Information 
of this paper.



\section{Optimization of difference-based estimators of autocovariance function}\label{sec:Optimization}

 In what follows we will say that an estimate $W_n$ is consistent if $\MSE[W_n] \to 0$ 
 as $n\to\infty$. 

\subsection{On the consistency of generalized difference-based estimators
for variance of $m$-dependent processes}~\label{sec.NoGoalTheo}

 The next theorem highlights that in a change-point segment regression with 
 zero mean, stationary, $m$-dependent errors, consistent estimation of the 
 variance $\gamma_0$ based on difference schemes already restricts this class to 
 estimators with gap $h$ at least $m+1$, see \eqref{eq.DBEk}.
 Some technical details are deferred to Appendix~\ref{sec.ProofsNoGoal}.
 
 \begin{theo}\label{theo.NoGoal}
  In the segment regression model \eqref{eq.NPR}-\eqref{eq.piecewise.function}-\eqref{eq.DistBetweenJumps}
  with zero mean, $m$-dependent, stationary errors, any consistent difference-based 
  estimator for the variance $\gamma_0$ given by \eqref{eq.DBEk}-\eqref{eq.a},
  has necessarily \emph{gap} at least $m+1$. More precisely, let $g$ and $m$
  be fixed integers with $g\geq m$ and assume that there exists an integer $N>1$ such that 
  $n=N(g+1)$. Then, in \eqref{eq.DBEk}, the vector of weights $\bs{w}_n$ must have the form:
  \[
    \bs{w}_n
    =
    ( \bs{v}_0 \quad \bs{v}_1 \quad \cdots \quad \bs{v}_{N-1} )^\top\in\R^n,
  \]
  where $\bs{v}_i = ( d_{i\,\cdot\,g} \quad 0 \quad \cdots \quad 0)^\top\in\R^{g+1}$,
  $i=0,\ldots,N-1$; $d_0\neq 0$, $d_{k\,\cdot\,g}\neq 0$ for some $1\leq k\leq N-1$, and 
  $\sum_{j=0}^{N-1}d_{j\,\cdot\,g}=0$. Here $j\cdot g$ denotes the multiplication of 
  the index $j$ by the integer $g$.
 \end{theo}
 \begin{pf}[Idea of proof]
   Since a consistent estimate is necessarily asymptotically unbiased, our line of
  argument consists of showing that for any difference-based estimate
  satisfying \eqref{eq.DBEk}-\eqref{eq.a} to be an asymptotically 
  unbiased estimate of $\gamma_0$, it is necessary that its 
  \emph{gap} be at least $m+1$. The technical details of the proof can be
  found in Appendix~\ref{sec.ProofsNoGoal}.  
 \end{pf}
 
 A key part of the proof of Theorem~\ref{theo.NoGoal} consists of
 computing the bias of a generalized difference-based estimator of order $l$ and 
 gap 1. This is provided by Lemma~\ref{lem.lim.DF.PC} below and
 in order to present it now we introduce some notation. 
 Recall that $f_i$ denotes $f(x_i)$ and for $i<n-l$, $f_{i:(i+l)}$ denotes the vector 
 $(f_i \quad f_{i+1} \quad \cdots \quad f_{i+l} )^\top\in\R^{l+1}$.
 Let
\begin{equation}\label{eq.Dtilde}
 \wt{D}
 =
  \begin{pmatrix}
  d_0 & d_1 & d_2 & \cdots &    d_l & 0\\
  0 &   d_0 & d_1 &    d_2 & \cdots & d_l
 \end{pmatrix}.
\end{equation}
  For $x\in\R^d$, $\|x\|$ denotes its Euclidean norm.

 \begin{lem}\label{lem.lim.DF.PC}
 Set $n_l := n-(l+2)>0$ and suppose that the conditions of Theorem~\ref{theo.NoGoal}
 hold. Let ${\bf Q}_1({\bf Y}, \bs{w}_l)$ be the difference-based estimator of 
 order $l$ and gap 1 given in \eqref{eq.DBEk}.
 Then for $l < 4(m+1)$:
\begin{equation}\label{eq.lem.BIAS.DBE}
  \BIAS[ {\bf Q}_1({\bf Y}, \bs{w}_l) ]
  =
  n_l^{-1}\,\mc O \left( 
  J_K\,\sum_{k=1}^l\left( \sum_{j = k}^l\, d_j \right)^2\right).
\end{equation}
\end{lem}
\begin{pf}[Idea of proof]
 Combining \eqref{eq.DBE.AsympEV} and Proposition~\ref{lem.trace.DS},
 cf.~Appendix~\ref{sec.ProofsNoGoal}, we get that
 \[
 \BIAS[ {\bf Q}_1({\bf Y}, \bs{w}_l) ]
 =
 n_l^{-1}\,\mc O \left( \sum_{j=1}^{n_l}\,\| \wt{D}f_{j:(j+l+1)} \|^2 \right).
 \]
 The right-hand side of this equation is given explicitly
 in the aforementioned Appendix.
\end{pf}

 Since the difference-based estimator of order $l=2$ and
 gap $(m+1)$, ${\bf Q}_{m+1}({\bf Y}, \bs{w}_2)$ cf.~\eqref{eq.DBEk}, can be represented 
 equivalently with the weights
 \[
 \bs{w}_{2(m+1)+1}
 =
  ( d_0 \quad \overbrace{0 \cdots 0}^{m \footnotesize{\mbox{ times}}}
  \quad d_1 \quad \overbrace{0 \cdots 0}^{m \footnotesize{\mbox{ times}} } \quad d_2),
 \]
 where $d_0+d_1+d_2 = 0$, Lemma~\ref{lem.lim.DF.PC} tells us
 that the bias of such estimator is of order $2(m+1)\max (d_0^2, d_2^2)\,J_k/n$. 
 Similar calculations show that for $l=3$, $\BIAS[ {\bf Q}_{m+1}({\bf Y}, \bs{w}_3) ]=\mc O ( 3(m+1)\,J_k/n )$.
 That is, considering a high-order difference-based estimator ($l\geq 2$)
 with gap $m+1$ will increase the bias by a magnitude of order 
 $l(m+1)$, hence we restrict to the case $l=2$ as our main focus
 is in bias minimizing estimators.

\subsection{Bias minimizer for autocovariance difference-based estimators 
of second order and gap $m+1$}~\label{sec.bias.minimizer}

  This section contains proofs for Theorems~\ref{theo.MSE} and \ref{theo.Bias.diff-based.MAq}.
  The auxiliary  
  Lemmas~\ref{lem.ExpValEtai2}, \ref{lem.ExpValEtai4}, \ref{lem.A}, \ref{lem.B}, 
  \ref{lem.MaxCorrProcesses}, \ref{lem.exp-val.B}, \ref{lem.exp-val.prod} and 
  \ref{lem.exp-val.Psi} can be found in Appendix~\ref{sec.Proofs.bias.minimizer}.
 \begin{pf}[Proof of Theorem~\ref{theo.MSE}]  
 Let $n_m = n - 2(m+1)$.
 In Eq.~\eqref{eq.gamma0} write $b_i(d) = \delta_i(d) + \eta_i(d)$,
 where $\delta_i(d)=f_i-f_{i+m+1}+d\,(f_{i+2(m+1)}-f_{i+m+1})$, 
 $\eta_i(d)=\varepsilon_i-\varepsilon_{i+m+1} + d(\varepsilon_{i+2(m+1)}-\varepsilon_{i+m+1})$,
 i.e., $\delta_i(d)$ \emph{carries} the signal $f$ while $\eta_i(d)$ carries
 the errors $\varepsilon$.
  
 Since $\ME [b_i^2(d)] = \delta_i^2(d) + \ME [\eta_i^2(d)]$, from Lemmas~\ref{lem.ExpValEtai2} 
 and \ref{lem.A} now follows that
  \begin{align}
  \ME[ \wh{\gamma}^{(m)}_0(d) ]
  =
  \gamma_0 + q_0(d)\,J_K/n_m,
  \quad 
  q_0(d) = \frac{(m+1)(d^2+1)}{2(d^2+d+1)}, \label{eq.ExpValGamma0}
  \end{align}
  recall that $J_K = \sum_{j=0}^{K-1}\,(a_{j+1}-a_j)^2$.
  
  Set $P(d) = 2(d^2+d+1)$. Since $\ME[b_i^4(d)] = \delta_i^4(d) + 6P(d)\,\delta_i^2(d)\gamma_0 + \ME[\eta_i^4(d)]$,
  we get from Lemmas~\ref{lem.ExpValEtai2} and \ref{lem.ExpValEtai4}:
  \begin{align*}
   \VAR( b_i^2(d) )
   &=
   4P(d)\gamma_0\,\delta_i(d) + 2P^2(d)\gamma_0^2\\
   \COV(b_i^2, b_j^2)
   &=
   4\delta_i(d)\delta_j(d) \ME[ \eta_i(d)\eta_j(d) ] + \ME [\eta_i^2(d)\eta_j^2(d)] - P^2(d)\gamma_0^2.
  \end{align*}
  
  Set ${\bf A} = 8\sum_{i,j}\,\delta_i(d)\delta_j(d) \ME[ \eta_i(d)\eta_j(d) ]$
  and ${\bf B} = 2\sum_{i,j}\,\ME [\eta_i^2(d)\eta_j^2(d)]$. Then
  \begin{align*}
   \VAR &(\wh{\gamma}_0^{(m)}(d))
   =
   \frac{n_m^{-2}}{P^2(d)}\left( 2n_mP^2(d)\gamma_0^2 
   + 
   4(m+1)(d^2+1)P(d)\gamma_0\,J_K
   +    
   {\bf A} + {\bf B} 
   - n_m^2P^2(d)\gamma_0^2\right)\\
   \intertext{which after an application of Lemmas~\ref{lem.A}-\ref{lem.B} from Appendix~\ref{sec.Proofs.bias.minimizer} 
   becomes}
   &=
   n_m^{-2}\left( p_1(d;\gamma_{(\cdot)})\gamma_0J_K 
   + 
   n_m p_2(d;\gamma_{(\cdot)})\gamma_0^2 
   + 
   p_3(d;\gamma_{(\cdot)}) \right),
  \end{align*}
  with
 \begin{align}
	p_1(d;\gamma_{(\cdot)}) 
	&=
	\frac{2(m+1)(d^4+1) 
	+ 2 \sum_{h=1}^m\,q_h^{(m)}(d)\,\rho_h}{(d^2+d+1)^2},\label{eq.mQ1}\\
	p_2(d;\gamma_{(\cdot)})
	&=
	\frac{ P^2(d) + \sum_{h=1}^{3m+2}\Lambda_h(d;\gamma_{(\cdot)})/\gamma_0^2 }{2(1+d+d^2)^2},\label{eq.mQ2}\\
	\intertext{and}
	p_3(d;\gamma)
	&=
	-\frac{1}{P^2(d)}\sum_{h=1}^{3m+2}\,h\,\Lambda_h(d;\gamma_{(\cdot)}) / \gamma_0^2, \label{eq.mQ3}
 \end{align}
 where $q_h^{(m)}(d) = [2(m+1) - 3h](d^4 + 1) + d^2 h$ and $\Lambda_h(d;\gamma_{(\cdot)})$
 is given in Lemma~\eqref{lem.B}, see Appendix~\ref{sec.Proofs.bias.minimizer}.
 
 Now we consider $\wh{\gamma}_h^{(m)}(d)=\wh{\gamma}_0^{(m)}(d) - \wh{\delta}^{(h)}$,
 cf.~\eqref{eq.gammak}.
 In light of the calculations above we only need to focus on 
 $\wh{\delta}^{(h)}$, cf.~\eqref{eq.odb.h}, $1\leq h\leq m < n$. The 
 arguments leading to $\ME[\wh{\gamma}^{(m)}_0(d)]$, cf.~\eqref{eq.ExpValGamma0},
 allow us to get that
\begin{equation}\label{eq.Exp-gammahat-h1}
  \ME[ \wh{\delta}^{(h)} ]
  =
  \gamma_0-\gamma_h + \frac{h\,J_K}{2(n-h)}.
\end{equation}
  Observe that a combination of ~\eqref{eq.ExpValGamma0}-\eqref{eq.Exp-gammahat-h1}
  yields
  \begin{equation}~\label{eq.Qast0}
  p_0^\ast(d)
  =
  p_0(d) - h/2.   
  \end{equation}

  Next, combine \eqref{eq.Exp-gammahat-h1}-\eqref{eq.Exp-gammahat2-h1}-\eqref{eq.Exp-Ah}
  and Lemma~\ref{lem.exp-val.B}, cf.~Appendix~\ref{sec.Proofs.bias.minimizer}, to get that 
 \begin{equation}\label{eq.VAR-gammahat-h1}
  \VAR( \wh{\delta}^{(h)} )
  =
  [ 2(n-h) ]^{-2}[ F_h( \gamma_{(\cdot)} )\,J_K + S_{1,n}^{(h)} ].
 \end{equation}
 The $\COV ( \wh{\gamma}_0^{(m)}(d) , \wh{\delta}^{(h)} )$ is given by
 \eqref{eq.exp-val.prod} (see Lemmas~\ref{lem.exp-val.prod} and \ref{lem.exp-val.Psi} 
 in the aforementioned Appendix). Arranging all these 
 terms we get that
 \begin{align}
  p_1^\ast(d;\gamma_{(\cdot)})
  &=
  p_1(d;\gamma_{(\cdot)}) + \frac{1}{4}
  \,F_h(\gamma_{(\cdot)}) - \frac{8(d^2-1)}{P(d)}\,V_h(\gamma_{(\cdot)}), \label{eq.Qast1}\\
  p_2^\ast(d;\gamma_{(\cdot)})
  &=
  p_2(d;\gamma_{(\cdot)}) + \frac{1}{4n}S_{1,n}^{(h)} + \frac{S_{2,n}^{(h)}(d)}{nP(d)}, \label{eq.Qast2}\\
  p_3^\ast(d;\gamma_{(\cdot)})
  &=
  p_3(d;\gamma_{(\cdot)})\gamma_0^2 + \frac{\gamma_0(\gamma_0-\gamma_h)}{P(d)}.~\label{eq.Qast3}
 \end{align}
 This concludes the proof.
 \end{pf}

  Observe that neither the functions $F_h(\cdot)$, $V_h(\cdot)$ or the quantities
 $S_{1,n}^{(h)}$, $S_{2,n}^{(h)}(d)$, cf.~\eqref{eq.Qast1}-\eqref{eq.Qast2}, depend 
 on the signal $f$ but only on the unknown autocovariance function $\gamma_{(\cdot)}$; also for any $1\leq h\leq m$,
 $S_{1,n}^{(h)}=\mc O(n)$ and $S_{2,n}^{(h)}(d)= \mc O(n)$, for any $d\in\R$.
  As a consequence of Theorem~\ref{theo.MSE} we deduce that if $J_K= o(\sqrt{n})$,
  then the estimates given by \eqref{eq.Delta(121).1}-\eqref{eq.Delta(121).2}
  are $\sqrt{n}$-consistent. Hence we obtain the following as an immediate
  consequence.
  
  
   \begin{theo}\label{prop.consistency}
  Suppose that the assumptions of Theorem~\ref{theo.MSE} hold.
  Let $\wh{\gamma}_h^{(m)}(d_{h,m})$ be given by \eqref{eq.Delta(121).1}-\eqref{eq.Delta(121).2}.
  If $J_K = o(\sqrt{n})$, then for $h = 0,\ldots,m$,
  \[
    \BIAS[ \wh{\gamma}_h^{(m)}(d_{h,m}) ] 
    =
    o (n^{-1/2})\quad \mbox{ and } \quad
    | \wh{\gamma}_h^{(m)}(d_{h,m}) - \gamma_h | 
    =
    \mc O_{\bf P}(n^{-1/2}).
  \]
 \end{theo}

 The remaining part of this section is devoted to the proof of 
 Theorem~\ref{theo.Bias.diff-based.MAq}.
 In a nutshell this theorem establishes that 
 if the error process in \eqref{eq.NPR} is Gaussian, $m$-dependent and 
 with autocovariance satisfying \eqref{eq.RegionMaximalCorrelation}
 then the extended bias of $\wh{\gamma}_0^{(m)}(d)$, cf.~\eqref{eq.ExtBias}, 
 is minimized at $d=1$. The same result holds when the correlation is 
 non-negative (see lemma below). 
 
 \begin{pf}[Proof of Theorem~\ref{theo.Bias.diff-based.MAq}]
  Since $\argmin_{d\in\R}p_0(d) = 1$, cf.~Theorem~\eqref{cor.BIASgamma0}, 
  we only need to show that
  \begin{equation}\label{eq.proff.theo.Bias.diff-based}
   \argmin_{d\in\R}p_1(d) = 1.
  \end{equation}
    
  For the $m$-dependent processes fulfilling {\bf 1.}, cf.~Theorem~\ref{theo.Bias.diff-based.MAq}, 
  it suffices to note that 
  $\rho\in(\max\{\,-1,-8/(3m^2)\,\},1]$, and the validity
  of \eqref{eq.proff.theo.Bias.diff-based} follows 
  by Lemma~\ref{lem.MaxCorrProcesses} of Appendix~\ref{sec.Proofs.bias.minimizer}. 
  Lemma~\ref{prop.OptBiasMA} below shows the validity of \eqref{eq.proff.theo.Bias.diff-based} for 
  the processes fulfilling {\bf 2.} The proof is complete.
 \end{pf}

  \begin{lem}\label{prop.OptBiasMA}
    Suppose that the assumptions of Theorem~\ref{theo.MSE} hold.
  Additionally, assume that the correlation function $\rho_h=\gamma_h/\gamma_0$
  is non-negative.
  Then $p_1(d;\gamma_{(\cdot)})$ and $\BIAS^\ast[\wh{\gamma}_0^{(m)}(d)]$
  are minimized at $d=1$.
 \end{lem}

 \begin{pf}%
  It is convenient to keep in mind \eqref{eq.mQ1}.
  For $m = 1$, apply Lemma~\ref{lem.MaxCorrProcesses} of Appendix~\ref{sec.Proofs.bias.minimizer}
  with $\rho=\rho_h\geq 0$. For $m>1$, since 
  $(d^4+1)/(d^2+d+1)^2\geq 2/9$ for all $d\in\R$, 
  it suffices to show that, 
  \begin{equation}\label{eq.lower.bound.sumUfunctions.general}
   \sum_{h=1}^m\,\rho_h\left[ q_h^{(m)}(d) - q_h^{(m)}(1) \right]\geq 0, \mbox{ for all }d\in\R.
  \end{equation}
  Let $\rho_{\min}=\min\,\{\rho_1,\ldots,\rho_m\}\geq0$. Observe that 
  \[
   \sum_{h=1}^m\,\rho_h\left[ q_h^{(m)}(d) - q_h^{(m)}(1) \right]
   \geq 
   \sum_{h=1}^m\,\rho_{\min}\left[ q_h^{(m)}(d) - q_h^{(m)}(1) \right]
   \geq0,
   \]
  the latter follows again from Lemma~\ref{lem.MaxCorrProcesses} with $\rho=\rho_{\min}$. 
  This shows the validity of \eqref{eq.lower.bound.sumUfunctions.general}.
 \end{pf}
 

\section{On the $\sqrt{n}$-consistency of difference-based autocovariance 
estimates in nonparametric H\"older regression with stationary $m$-dependent 
errors}~\label{sec:HolderRegression}

 Thus far we have conducted a non-asymptotic analysis of bias and variance
 of the class of estimates \eqref{eq.gamma0}-\eqref{eq.gammak} for Gaussian
 errors along with the condition \eqref{eq.DistBetweenJumps}. This 
 led us to the bias-reducing estimates $\wh{\gamma}^{(m)}_h(d_{h,m})$, 
 cf.~Eq.~\eqref{eq.Delta(121).1}-\eqref{eq.Delta(121).2}. Moreover,
 we showed that $J_K = o(\sqrt{n})$ is a sufficient condition for the 
 $\sqrt{n}$-consistency of $\wh{\gamma}_h^{(m)}(d_{h,m})$ in the 
 segment regression model~\eqref{eq.NPR}-\eqref{eq.piecewise.function}-\eqref{eq.DistBetweenJumps},
 see Theorem~\ref{prop.consistency} above.
 In this section,
 we show that $\wh{\gamma}^{(m)}_h(d_{h,m})$, 
 are $\sqrt{n}$-consistent estimates of $\gamma_h$, $h=0,\ldots,m$ in a general 
 regression model with possibly non-Gaussian, stationary, $m$-dependent errors.
 
 More precisely, we consider observations from the regression model introduced
 in \eqref{eq.NPR} when the unknown signal $f$ admits 
 the representation
\begin{equation}\label{eq.piecewise.Lip.function}
 f(x) = \sum_{j=0}^{K_n-1}\,a_j(x)\ind_{[\tau_j,\tau_{j+1})}(x),\quad x\in[0,1),
\end{equation}
where $a_j:[0,1)\to\R$ are unknown H\"older functions, i.e.,
for all $x,y\in[0,1)$ there exists a generic constant $C> 0$ and an index $\alpha_j\geq 0$ such that
\begin{equation}\label{eq.Holder}
 | a_j(x) - a_j(y) | \leq C\,| x-y |^{\alpha_j},\quad j=0,\ldots,K_n-1.
\end{equation}
  We will assume that $f$ has effectively $K_n$ discontinuity points. Namely, 
  let $\vt_j:=a_j(\tau_{j+1}^{-})-a_{j+1}(\tau_{j+1})$
  be the $j$-th jump of $f$, we will assume that there exists a number $c>0$ such 
  that $|\vt_j|>c$ for $j=1,\ldots,K_n-1$. 
  Again, the change-points $0=\tau_0<\tau_1<\cdots<\tau_{K_n-1}<\tau_{K_n}=1$ and
  $K_n$ are unknown and may depend on $n$ now. 
  We assume that the errors are a sample from a zero 
  mean $m$-dependent process with stationary moments up to order 4,
  $(\varepsilon_i)_{ i\geq 1}$.
  The main result of this section is now stated.

 \begin{theo}\label{theo.AsympProperties.Gammah}
  Consider the model given by Eqs.~\eqref{eq.NPR}-\eqref{eq.piecewise.Lip.function}-\eqref{eq.Holder}.
  Suppose that Eq.~\eqref{eq.DistBetweenJumps} holds with $K$ replaced by
  $K_n$. Assume that $\vt^\ast:=\max_j\{ | \vt_j |  \}<\infty$.
  Let $\wh{\gamma}_h^{(m)}(d_{h,m})$ be given
  by Eqs.~\eqref{eq.Delta(121).1}-\eqref{eq.Delta(121).2}.
  Then, for $h = 0,\ldots,m$,
  \begin{align}
    \BIAS[ \wh{\gamma}_h^{(m)}(d_{h,m}) ] 
    &= 
    \mc O\left( \frac{K_n}{{n}} + \sum_{j=1}^{K_n}\,n^{-2(\alpha_j+1/2)} \right). \label{eq.AsympBIASgammah}
    \intertext{If additionally, $K_n= o(\sqrt{n})$,}
   | \wh{\gamma}_h^{(m)}(d_{h,m}) - \gamma_h | 
   &= 
   \mc O_{\bf P}(n^{-1/2}).~\label{eq.Rootn.gammah}    
  \end{align}
 \end{theo}
 \begin{pf}
  The auxiliary results to this proof can be found in Appendix~\ref{sec.AppendixIII}.
  The validity of Eq.~\eqref{eq.AsympBIASgammah} follows from Lemma~\ref{lem.BIAS.Holder}.
  Since $\alpha_\ast:=\min_j\{\alpha_j\} \geq 0$, combining Lemmas~\ref{lem.VAR.Holder} 
  and \ref{lem.COV.Holder} we get that
 \begin{align*}
  \ME[ \sqrt{n}( \wh{\gamma}_h^{(m)}(d_{h,m}) - \gamma_h ) ] 
  &= 
  \mc O( K_n / \sqrt{n} )\\
  \VAR( \sqrt{n}( \wh{\gamma}_h^{(m)}(d_{h,m}) - \gamma_h ) )
  &=
  \mc O( K_n^2/n ) ) + \mc O(1).
 \end{align*}
 The validity of Eq.~\eqref{eq.Rootn.gammah} now follows by using $K_n = o(\sqrt{n})$.
\end{pf}


\section{Projection-based covariance matrix estimation}\label{sec:CovMatrix}
  In general, autocovariance estimates based on difference schemes may 
  lead to a non-positive definite (ill-defined) covariance matrix estimate 
  in model \eqref{eq.NPR}. In order to overcome this problem in this section we 
  propose a projection-based covariance estimation method. 

Assume that the vector $\bs{Y} = ( y_1 \quad \cdots \quad y_n)\in\R^n$ follows
the model~\eqref{eq.NPR} with signal $f\in\mc{F}$ (some class of functions) and zero mean, stationary 
$m$-dependent errors $\bs{\varepsilon}=(\varepsilon_i)_{1\leq i\leq n}$,
and associated covariance matrix
\begin{equation}\label{eq.GammaMatrix}
 \Gamma^{(m)}
 =
 {\tiny{
 \begin{pmatrix}
 \gamma_0 & \gamma_1 & \cdots & \gamma_m &&\\
 \gamma_1 & \gamma_0 &        &          &&\\
   \vdots &   \ddots & \ddots &          & \ddots &0&\\
   \gamma_m &&&&&\\
   &\ddots& & &\ddots &&\gamma_m\\
   &0&&&&& \vdots\\
   &&&&& \gamma_0 & \gamma_1\\
   &&& \gamma_m & \cdots & \gamma_1 & \gamma_0
 \end{pmatrix}}_{n\times n}.
 }
\end{equation}
That is, $\Gamma^{(m)}\in\mc{C}_n^{(m)}$, the set of all the $n\times n$ symmetric, positive semidefinite
$(m+1)$-banded Toeplitz matrices. Observe that $\mc{C}_n^{(m)}$ is a subset of $\msc{H}$, the Hilbert space of all the
$n\times n$ symmetric matrices with inner product $\inner{A}{B}=\mbox{tr}( A^\top\,B)$
and induced (Frobenius) norm $\|A\|_F=\sum_{i,j=1}^n\,a_{i,j}^2$, $A,B\in \msc{H}$.
Let $\mc{S}_n$ and $\mc{T}_n^{(m)}$ denote the subsets of $\msc{H}$ of all the
positive semidefinite and $(m+1)$-banded Toeplitz matrices, respectively. 
Note that $\mc{C}_n^{(m)}=\mc{S}_n\cap\mc{T}_n^{(m)}$.
We define $\MSE_{(f,\Gamma^{(m)})}[\wh{\Gamma}] := \ME_{\Gamma^{(m)}}\| \Gamma^{(m)} - \wh{\Gamma} \|^2_F$, 
where $f\in \mc{F}$, $\Gamma^{(m)}\in \mc{C}_n^{(m)}$ is given by \eqref{eq.GammaMatrix},
and $\wh{\Gamma}\in \msc{H}$, is some estimator. The following lemma is 
essential to get the main result of this section.

\begin{lem}\label{lem.RiskReducingEstimate}
 Let $\Theta$ be a closed convex set of a Hilbert space $\msc{H}$ with
 inner product $\inner{\cdot}{\cdot}$ and induced norm $\|\vt\|=\inner{\vt}{\vt}^{1/2}$.
 Let $Y\sim \prob_{\vt_0}$ where $\vt_0\in \Theta$ and let $\wh{\vt}\in \msc{H}$
 be an estimate of $\vt_0$. Let $P_\Theta(\wh{\vt}) = \argmin \{ \| \wh{\vt} - \vt \|: \vt\in \Theta \}$
 be the unique projection of $\wh{\vt}$ onto $\Theta$.
 Then, 
 \begin{equation}\label{eq.RiskReducingProperty}
  \ME_{\vt_0}  \| \vt_0 - P_\Theta(\wh{\vt}) \|^2 
  \leq 
  \ME_{\vt_0} \| \vt_0 - \wh{\vt} \|^2.
 \end{equation}
\end{lem}
\begin{pf}
The well-known projection theorem, cf.~\cite{Luenberger.68}, p.~69, 
characterizes $P_\Theta(\wh{\vt})$ by the condition that
\begin{equation}\label{eq.ProjPrinciple}
  \inner{\wh{\vt}-P_\Theta(\wh{\vt})}{\vt-P_\Theta(\wh{\vt})} \leq 0, \quad \forall \vt\in \Theta. 
\end{equation}
Then observe that
\begin{equation}\label{eq.ProjPrincipleAux}
 \|\wh{\vt}-\vt_0\|^2
 =
 \| P_\Theta(\wh{\vt}) - \vt_0 \|^2 + \| (I-P_\Theta)(\wh{\vt}) \|^2
 -2\inner{ \wh{\vt}-P_\Theta(\wh{\vt}) }{ \vt_0 - P_\Theta(\wh{\vt}) }
 \geq
 \| P_\Theta(\wh{\vt}) - \vt_0 \|^2.
\end{equation}
 The latter follows by \eqref{eq.ProjPrinciple} since $\vt_0\in \Theta$. 
 The result follows by taking expectations in \eqref{eq.ProjPrincipleAux}.
\end{pf}

This general principle of convex optimization allows us to get well-defined 
covariance estimates with reduced risk by properly projecting preliminary 
(and possibly ill-defined) estimates onto $\mc{C}_n^{(m)}$. More precisely:

\begin{theo}\label{theo.CovarianceEstimate}
 Let $\left(\wh{\gamma}_0(\bs{Y}) \quad \cdots \quad \wh{\gamma}_m(\bs{Y})\right)$
 be any estimate of the vector $\left({\gamma}_0 \quad \cdots \quad {\gamma}_m\right)$
 whose corresponding matrix is denoted by $\wh{\Gamma}$ and has the form \eqref{eq.GammaMatrix}.
 Let us define $\wh{\Gamma}^\ast := P_{\mc{C}_{n}}(\wh{\Gamma})=\argmin \{ \| \wh{\Gamma} - \Gamma^{(m)} \|_F: 
 \Gamma^{(m)}\in \mc{C}_{n}^{(m)} \}$, i.e., the unique projection of $\wh{\Gamma}$ onto $\mc{C}_n^{(m)}$ 
 w.r.t.~$\|\cdot\|_F$.
 Then, 
\[
 \MSE_{(f,\Gamma^{(m)})}[ \wh{\Gamma}^\ast ] \leq \MSE_{(f,\Gamma^{(m)})}[ \wh{\Gamma} ],
\mbox{ for all }(f,\Gamma^{(m)})\in \mc{F}\times \mc{C}_n^{(m)}.
\]
 \end{theo}
 \begin{pf}
  Since $\mc{C}_n$ is the intersection of ($\mc{S}_n$ and $\mc{T}_n^{(m)}$) two closed 
  convex sets of $\msc{H}$, the result follows by an application of Lemma~\ref{lem.RiskReducingEstimate}.
 \end{pf}

 \begin{rmk}
  The validity of this result does not depend on any distributional
 assumption about the errors $\bs{\varepsilon}$ neither on any specific form for 
 the family of signals $\mc{F}$.
 Observe also that Theorem~\ref{theo.CovarianceEstimate} holds true for
 any $b$-banded Toeplitz matrix for $1\leq b \leq n-1$ provided $b$ is fixed.
 Now, we show how to compute $\wh{\Gamma}^\ast$, that is,
 the nearest symmetric, positive semidefinite
 banded Toeplitz matrix to a given covariance matrix estimate.
 \end{rmk}

 \vspace{0.5cm}
 \subsection{Alternating projections method}~\label{sec:AlternatingAlgorithm}
 
 In this subsection we utilize the notation introduced in Theorem~\ref{theo.CovarianceEstimate}.
 The representation of $\mc{C}_n^{(m)}$ as 
 the intersection of $\mc{S}_n$ and $\mc{T}_n^{(m)}$ suggests an alternating
 projection algorithm for the computation of $\wh{\Gamma}^\ast$: in order 
 to compute $\wh{\Gamma}^\ast$ we have to project iteratively onto $\mc S_n$
 and then onto $\mc T_n^{(m)}$ by repeating the operation
 \begin{equation}\label{eq.ProjectionAlgorithm}
  \wh{\Gamma} \leftarrow P_{\mc{T}_n^{(m)}}( P_{\mc{S}_n}(\wh{\Gamma}) ).
 \end{equation}

 Let $\lambda_i$ be the $i$-th eigenvalue of $\wh{\Gamma}$.
 The spectral decomposition of $\wh{\Gamma}=Q\,D\,Q^\top$, where 
 $D=\mbox{diag}(\lambda_i)$ and $Q$ is an orthogonal matrix containing the
 eigenvectors of $\wh{\Gamma}$ gives us
 \[
  P_{\mc{S}_n}(\wh{\Gamma}) = Q\,\mbox{diag}( \max( \lambda_i, 0 ) )\,Q^\top,
 \]
 see e.g., Theorem~3.2 of \cite{Higham.02}.
 
 It is well-known that $P_{\mc{T}_n^{(m)}}(\wh{\Gamma})$, the orthogonal projection of 
 $\wh{\Gamma}$ onto $\mc{T}_n^{(m)}$, is given by the $n\times n$ symmetric, banded 
 Toeplitz matrix whose first row is given by
 \begin{align*}
  t_k
  &=
  \frac{1}{n-k}\sum_{i=1}^{n-k}\,\wh{\gamma}_{i,i+k}, \quad k=0,\ldots,n-1,
 \end{align*}
 see e.g., Eqs.~(2.3)-(2.5) of \cite{Grigoriadis.etal.94}.
 
 Since $\mc{S}_n$ is not a linear subspace, 
 the alternating projection algorithm \eqref{eq.ProjectionAlgorithm} requires 
 a modification for it to converge.
 Such a modification is due to \cite{Dykstra.83} which combines a 
 beneficial correction to each projection which can be seen as a normal vector
 (in a geometric sense) to the corresponding convex set. 
 
\begin{alg}\label{alg.1}
 Given a symmetric matrix $S_0\in\R^{n\times n}$ this algorithm computes the
 \emph{nearest} Toeplitz covariance matrix to $S_0$ in the Frobenius norm.
 \begin{align*}
  DC_0 &= 0, P_0=S_0\\
  \mbox{for } k &=1,2,\ldots\\
  R_k &= P_{k-1}-DC_{k-1}, \% DC_{k-1}\mbox{ is Dykstra's correction.}\\
  X_k &= P_{\mc{S}_n}(R_k)\\
  DC_k &= X_k - R_k\\
  P_k &= P_{\mc{T}_n^{(m)}}( X_k )\\
  \mbox{end }
 \end{align*}
\end{alg}
  
 \vspace{-.25cm}
 The function \texttt{nearPDToeplitz}, in the \texttt{R} package \texttt{dbacf},
 implements this algorithm.
 According to Theorem~2 of \cite{Boyle.Dykstra.86} the sequence $P_k$, $k=0,1,\ldots,$
 converges to $P_{\mc{C}_n^{(m)}}(S_0)$, the orthogonal projection of the initial point $S_0$
 onto the closed convex set of symmetric, positive semidefinite, $(m+1)$-banded 
 Toeplitz matrices.

\section{Simulations}~\label{sec.Simulations}

  This section contains 2 simulation studies.
  The first one assesses the performance of \eqref{eq.Delta(121).1}-\eqref{eq.Delta(121).2}
  for autocorrelation estimation and their robustness against non-normally distributed 
  errors.
  In the second study we assess the performance of these estimates when the signal
  is a smooth function.

  In our simulations for the noise we consider a 1-dependent error model: $\varepsilon_i = r_0\delta_i + r_1\delta_{i-1}$
where $\delta_i$'s are i.i.~$\mc D$-distributed, $r_0 = [ \sqrt{1+2\gamma_1} +  \sqrt{1-2\gamma_1}]/2$
and $r_1 = [ \sqrt{1+2\gamma_1} -  \sqrt{1-2\gamma_1}]/2$ for $-1/2\leq \gamma_1 \leq 1/2$;
we will utilize $\mc D = \mc N(0,1)$ and to assess robustness against non-normality we 
will use $\mc D = t_4$. Since the autocorrelation of $\varepsilon_i$
at lag 1 satisfies $\rho_1 = \gamma_1$, we will assess estimation
of $\rho_1$ and $\rho_2=0$.

In this section and in the upcoming Section~\ref{sec.Applications} we will use $\wh{\rho}_O$ to denote autocorrelation estimates
based on the autocovariance bias-optimized estimates \eqref{eq.Delta(121).1}-\eqref{eq.Delta(121).2},
the symbols $\wh{\rho}_H$, $\wh{\rho}_M$, $\wh{\rho}_R$, $\wh{\rho}_P$, 
$\wh{\rho}_{HV}$ will denote \cite{Herrmann.92}'s, \cite{Muller.Stadtmuller.88}'s, 
adapted-\cite{Rice.84}'s (set $d=0$ in \eqref{eq.Delta(121).1}),
\cite{Park.etal.06} and \cite{Hall.VanKeilegom.03}'s autocorrelation estimates, respectively.
\cite{Hall.VanKeilegom.03}'s estimators depend on two smoothing parameters,
$l_1$ and $l_2$, following Section~3 of their paper we have chosen $l_1 = n^{0.4}$
and $l_2=\sqrt{n}$ in our simulations. Recall that except for $\wh{\rho}_{HV}$, the other
difference-based estimates require $m$ as a parameter, we have used $m=2$ in each
case. 

\subsection{Piecewise constant signal with 1-dependent errors}~\label{sec:ParkErrors_Chakarsignal}

The simulated signal in this study is an adaptation of that of \cite{Chakar.etal.16}.
More precisely, the signal is a piecewise constant function, $f$, with 6 change-points located 
at fractions $1/6 \pm 1/36$,
$3/6 \pm 2/36$, $5/6 \pm 3/36$ of the sample size $n$. In the first segment $f = 0$,
in the second one $f=10$, in the remaining segments $f$ alternates between 0 and 1,
starting with $f=0$ in the third segment. 


Tables~\ref{tab.mse_dbeMethods_ParksErrors_ChakarSignal}-\ref{tab.mse_dbeMethods_tdistErrors_ChakarSignal}
summarize the results for $n=1\,600$ obtained from 500 samples for Gaussian and $t_4$ errors,
respectively. Each table displays the $\MSE$ of $\wh{\rho}_O$, $\wh{\rho}_H$, $\wh{\rho}_M$, 
$\wh{\rho}_R$ and $\wh{\rho}_{HV}$. 
Table~\ref{tab.mse_dbeMethods_ParksErrors_ChakarSignal} shows that $\wh{\rho}_O$
overperfoms all the other estimates under comparison for all the values of $\gamma_1$;
note that for positive values of $\gamma_1$, $\wh{\rho}_O$ beats $\wh{\rho}_M$, 
$\wh{\rho}_R$ and $\wh{\rho}_{HV}$ up to 2 orders of magnitude.
Since $\wh{\rho}_O$ and $\wh{\rho}_H$ use the same statistic to estimate $\gamma_0$, 
it is expected that $\wh{\rho}_O$ only overperfoms $\wh{\rho}_H$ marginally. 
Similar results are obtained when the errors are heavy-tailed ($t_4$-distributed),
see Table~\ref{tab.mse_dbeMethods_tdistErrors_ChakarSignal}.
That is, the performance of $\wh{\rho}_O$ does not seem to depend on the Gaussianity
of the errors, as also suggested by Theorem~\ref{theo.AsympProperties.Gammah}.\medskip

\begin{table}[ht]
 \centering
 \caption{\footnotesize The $\MSE$ of various autocorrelation estimators of
 $\rho_1 = \gamma_1$ and $\rho_2 = 0$ under the 1-dependent error model $\varepsilon_i = r_0\delta_i + r_1\delta_{i-1}$
  where $\delta_i$'s are i.i.d.~$\mc N(0,1)$, $r_0 = [ \sqrt{1+2\gamma_1} +  \sqrt{1-2\gamma_1}]/2$
  and $r_1 = [ \sqrt{1+2\gamma_1} -  \sqrt{1-2\gamma_1}]/2$, based on 500 pseudo-samples
  of size 200. 
  Signal is specified in Section~\ref{sec:ParkErrors_Chakarsignal}.}~\label{tab.mse_dbeMethods_ParksErrors_ChakarSignal}
  \scalebox{.75}{
 \begin{tabular}{c|c|>{\columncolor[gray]{0.8}}c|c|c|c|c|c}
 \toprule[1.25pt]
  \hline
     && $\wh{\rho}_O$ & $\wh{\rho}_H$ &$\wh{\rho}_M$ & $\wh{\rho}_R$ & $\wh{\rho}_{HV}$ &\\              
  \hline
     & $\gamma_1=-0.5$ &  &  &  &  &\\              
  &$\rho_1 = \gamma_1$ & 0.0147 & 0.0162 & 0.0357 & 0.0359 & 0.8749  &\\ 
  &$\rho_2 = 0$ & 0.0035 & 0.0037 & 0.0049 & 0.0049 & 0.3520 &\\ 
     & $\gamma_1=-0.4$ & &  &  &  &\\              
  &$\rho_1 = \gamma_1$ & 0.0132 & 0.0144 & 0.0311 & 0.0313 & 0.7612 &\\ 
  &$\rho_2 =0$ & 0.0033 & 0.0036 & 0.0048 & 0.0049 & 0.3522 &\\ 
       & $\gamma_1=-0.2$ & &  &  &  &  &\\              
  &$\rho_1 = \gamma_1$ & 0.0077 & 0.0089 & 0.0198 & 0.0199 & 0.5526 &\\ 
  &$\rho_2 = 0$ & 0.0023 & 0.0024 & 0.0036 & 0.0037 & 0.3499 &\\ 
       & $\gamma_1=0$ & &  &  &  &\\              
  &$\rho_1 = \gamma_1$ & 0.0049 & 0.0057 & 0.0126 & 0.0126 & 0.3801 &\\ 
  &$\rho_2 = 0$ & 0.0016 & 0.0022 & 0.0038 & 0.0038 & 0.3520 &\\ 
       & $\gamma_1=0.2$ & &  &  &  &\\              
  &$\rho_1 = \gamma_1$ & 0.0023 & 0.0028 & 0.0062 & 0.0062 & 0.2377  &\\ 
  &$\rho_2 = 0$ & 0.0013 & 0.0018 & 0.0034 & 0.0034 & 0.3505 &\\ 
       & $\gamma_1=0.4$ & &  &  &  &\\              
  &$\rho_1 = \gamma_1$ & 0.0008 & 0.0010 & 0.0022 & 0.0022 & 0.1296 & \\ 
  &$\rho_2 = 0$ & 0.0010 & 0.0015 & 0.0033 & 0.0033 & 0.3512 & \\ 
       & $\gamma_1=0.5$ & &  &  &  &  &\\              
  &$\rho_1 = \gamma_1$ & 0.0006 & 0.0007 & 0.0011 & 0.0011 & 0.0885 & \\ 
  &$\rho_2 = 0$ & 0.0011 & 0.0018 & 0.0036 & 0.0036 & 0.3538 & \\ 
  \hline  
\bottomrule[1.25pt]
 \end{tabular}}
  \vspace{0.25cm}
  \caption*{\footnotesize Notation: 
  $\wh{\gamma}_O$, $\wh{\gamma}_H$, $\wh{\gamma}_M$, $\wh{\gamma}_R$, $\wh{\gamma}_{HV}$
  denote optimized difference-based, \cite{Herrmann.92}'s, \cite{Muller.Stadtmuller.88}'s,
  adapted-\cite{Rice.84}'s and \cite{Hall.VanKeilegom.03}'s estimates, 
  respectively.}
\end{table}

\begin{table}[ht]
 \centering
 \caption{\footnotesize The $\MSE$ of various autocorrelation estimators of
 $\rho_1 = \gamma_1$ and $\rho_2 = 0$ under the 1-dependent error model $\varepsilon_i = r_0\delta_i + r_1\delta_{i-1}$
  where $\delta_i$'s are i.i.d.~$t_4$, $r_0 = [ \sqrt{1+2\gamma_1} +  \sqrt{1-2\gamma_1}]/2$
  and $r_1 = [ \sqrt{1+2\gamma_1} -  \sqrt{1-2\gamma_1}]/2$, based on 500 pseudo-samples
  of size $n=1600$.
  Signal is specified in Section~\ref{sec:ParkErrors_Chakarsignal}.}~\label{tab.mse_dbeMethods_tdistErrors_ChakarSignal}
  \scalebox{.75}{
 \begin{tabular}{c|c|c|c|c|c|c|c}
 \toprule[1.25pt]
  \hline
     && $\wh{\rho}_O$ & $\wh{\rho}_H$ &$\wh{\rho}_M$ & $\wh{\rho}_R$ & $\wh{\rho}_{HV}$ &\\              
  \hline
     & $\gamma_1=-0.5$ &  &  &  &  &\\              
  &$\rho_1 = \gamma_1$ & 0.0069 & 0.0063 & 0.0119 & 0.0120 & 0.4620  &\\ 
  &$\rho_2 = 0$ & 0.0038 & 0.0037 & 0.0032 & 0.0033 & 0.1863 &\\ 
     & $\gamma_1=-0.4$ & &  &  &  &\\              
  &$\rho_1 = \gamma_1$ & 0.0059 & 0.0056 & 0.0104 & 0.0104 & 0.4042 &\\ 
  &$\rho_2 =0$ & 0.0035 & 0.0032 & 0.0031 & 0.0031 & 0.1868 &\\ 
       & $\gamma_1=-0.2$ & &  &  &  &  &\\              
  &$\rho_1 = \gamma_1$ & 0.0042 & 0.0042 & 0.0074 & 0.0075 & 0.2968 &\\ 
  &$\rho_2 = 0$ & 0.0023 & 0.0021 & 0.0023 & 0.0023 & 0.1879 &\\ 
       & $\gamma_1=0$ & &  &  &  &\\              
  &$\rho_1 = \gamma_1$ & 0.0035 & 0.0035 & 0.0047 & 0.0047 & 0.2018 &\\ 
  &$\rho_2 = 0$ & 0.0022 & 0.0018 & 0.0021 & 0.0021 & 0.1870 &\\ 
       & $\gamma_1=0.2$ & &  &  &  &\\              
  &$\rho_1 = \gamma_1$ & 0.0015 & 0.0016 & 0.0025 & 0.0025 & 0.1272 &\\ 
  &$\rho_2 = 0$ & 0.0014 & 0.0013 & 0.0016 & 0.0016 & 0.1873 &\\ 
       & $\gamma_1=0.4$ & &  &  &  &\\              
  &$\rho_1 = \gamma_1$ & 0.0008 & 0.0009 & 0.0010 & 0.0011 & 0.0690 & \\ 
  &$\rho_2 = 0$ & 0.0011 & 0.0010 & 0.0014 & 0.0014 & 0.1870 & \\ 
       & $\gamma_1=0.5$ & &  &  &  &  &\\              
  &$\rho_1 = \gamma_1$ & 0.0005 & 0.0006 & 0.0006 & 0.0006 & 0.0463 & \\ 
  &$\rho_2 = 0$ & 0.0010 & 0.0011 & 0.0014 & 0.0014 & 0.1849 & \\ 
  \hline  
\bottomrule[1.25pt]
 \end{tabular}}
  \vspace{0.25cm}
  \caption*{\footnotesize Notation: 
  $\wh{\gamma}_O$, $\wh{\gamma}_H$, $\wh{\gamma}_M$, $\wh{\gamma}_R$, $\wh{\gamma}_{HV}$
  denote optimized difference-based, \cite{Herrmann.92}'s, \cite{Muller.Stadtmuller.88}'s,
  adapted-\cite{Rice.84}'s and \cite{Hall.VanKeilegom.03}'s estimates, 
  respectively.}
\end{table}

\subsection{Smooth signal with 1-dependent errors}~\label{sec:ParkSetup}

The purpose of this study is to assess the performance of $\wh{\rho}_O$ 
for smooth signals when the bias of our estimators becomes less influential.
To this end, and following 
\cite{Park.etal.06}, we consider the mean function $f(x) = 300 x^3 (1 - x)^3\,\ind_{[0,1]}(x)$ 
\emph{sampled} at points $x_i = i/n$. 
As errors we use the 1-dependent process $(\varepsilon_i)$ introduced above
with $\mc D = \mc N(0,1)$, see Section~\ref{sec:ParkErrors_Chakarsignal}.

Park's estimation method consists of two stages. First, an optimized 
bimodal kernel method pre-filters the signal, the resulting residuals are then 
used to estimate the correlation structure via 
an ordinary difference-based method. Since our example follows 
\cite{Park.etal.06}' specifications, we use their reported results
and we apply the difference-based methods abovementioned for autocorrelation 
estimation of $\rho_1$ and $\rho_2 = 0$.

Table~\ref{tab.mse.Parks} summarizes the results obtained from 500 samples
of size $n=200$. 
According to this table, $\wh{\rho}_P$ and $\wh{\rho}_{HV}$ are
almost identical and these two estimates overperfom all the others in almost
all the cases. When the correlation is non-negative, $\wh{\rho}_{HV}$ is only marginally
better than $\wh{\rho}_O$, $\wh{\rho}_{M}$ and $\wh{\rho}_{R}$.
We conclude that for smooth signals, autocovariance estimates based only on difference schemes
($\wh{\rho}_M$, $\wh{\rho}_R$, $\wh{\rho}_{HV}$) are comparably as accurate as other estimates 
which rely on kernel-based residuals, for highly positively correlated noise 
($\gamma_1 = 0.5$) even the bias-optimized estimate $\wh{\rho}_O$ outperforms 
such estimates.

\begin{table}[ht]
 \centering
 \caption{\footnotesize The $\MSE$ of various autocorrelation estimators of
 $\rho_1 = \gamma_1$ and $\rho_2 = 0$ under the 1-dependent error model $\varepsilon_i = r_0\delta_i + r_1\delta_{i-1}$
  where $\delta_i$'s are i.i.d.~$\mc N(0,1)$, $r_0 = [ \sqrt{1+2\gamma_1} +  \sqrt{1-2\gamma_1}]/2$
  and $r_1 = [ \sqrt{1+2\gamma_1} -  \sqrt{1-2\gamma_1}]/2$, based on 500 pseudo-samples
  of size 200. Signal is specified in Section~\ref{sec:ParkSetup}.}~\label{tab.mse.Parks}
  \scalebox{.75}{
 \begin{tabular}{c|c|c|c|c|c|c|c|c}
 \toprule[1.25pt]
  \hline
     && $\wh{\rho}_O$ & $\wh{\rho}_H$ &$\wh{\rho}_M$ & $\wh{\rho}_R$ & $\wh{\rho}_P$ & $\wh{\rho}_{HV}$ &\\              
  \hline
     & $\gamma_1=-0.5$ &  &  &  &  &  &\\              
  &$\rho_1 = \gamma_1$ & 0.0318 & 0.0210 & 0.0161 & 0.0168 & {\bf 0.0029} & 0.0071 &\\ 
  &$\rho_2 = 0$ & 0.0392 & 0.0314 & 0.0264 & 0.0263 & {\bf 0.0076} & 0.0077 &\\ 
     & $\gamma_1=-0.4$ & &  &  &  &  &\\              
  &$\rho_1 = \gamma_1$ & 0.0268 & 0.0182 & 0.0141 & 0.0147 & {\bf 0.0038} & 0.0073 &\\ 
  &$\rho_2 =0$ & 0.0314 & 0.0264 & 0.0205 & 0.0207 & {\bf 0.0065} & 0.0073 &\\ 
       & $\gamma_1=-0.2$ & &  &  &  &  &\\              
  &$\rho_1 = \gamma_1$ & 0.0220 & 0.0171 & 0.0133 & 0.0133 & {\bf 0.0062} & 0.0084 &\\ 
  &$\rho_2 = 0$ & 0.0221 & 0.0181 & 0.0144 & 0.0145 & {\bf 0.0060} & 0.0066 &\\ 
       & $\gamma_1=0$ & &  &  &  &  &  \\              
  &$\rho_1 = \gamma_1$ & 0.0156 & 0.0128 & 0.0097 & 0.0098 & {\bf 0.0064} & 0.0069 &\\ 
  &$\rho_2 = 0$ & 0.0164 & 0.0133 & 0.0103 & 0.0104 & {\bf 0.0067} & {\bf 0.0067} &\\ 
       & $\gamma_1=0.2$ & &  &  &  &  &  \\              
  &$\rho_1 = \gamma_1$ & 0.0125 & 0.0112 & 0.0083 & 0.0082 & 0.0065 & {\bf 0.0054} &\\ 
  &$\rho_2 = 0$ & 0.0122 & 0.0094 & {\bf 0.0074} & {\bf 0.0074} & 0.0081 & 0.0080 &\\ 
       & $\gamma_1=0.4$ & &  &  &  &  &  \\              
  &$\rho_1 = \gamma_1$ & 0.0067 & 0.0064 & 0.0043 & 0.0044 & 0.0049 & {\bf 0.0035} & \\ 
  &$\rho_2 = 0$ & 0.0089 & 0.0070 & {\bf 0.0048} & 0.0051 & 0.0109 & 0.0091 & \\ 
       & $\gamma_1=0.5$ & &  &  &  &  &  \\              
  &$\rho_1 = \gamma_1$ & 0.0051 & 0.0052 & 0.0034 & 0.0035 & 0.0589 & {\bf 0.0029} & \\ 
  &$\rho_2 = 0$ & 0.0097 & 0.0079 & {\bf 0.0054} & 0.0058 & 0.1059 & 0.0096 & \\ 
  \hline  
\bottomrule[1.25pt]
 \end{tabular}}
  \vspace{0.25cm}
  \caption*{\footnotesize Notation: 
  $\wh{\rho}_O$, $\wh{\rho}_H$, $\wh{\rho}_M$, $\wh{\rho}_R$, $\wh{\rho}_P$, $\wh{\rho}_{HV}$
  denote optimized difference-based, \cite{Herrmann.92}'s, 
  \cite{Muller.Stadtmuller.88}'s, adapted-\cite{Rice.84}'s, \cite{Park.etal.06}'s and 
  \cite{Hall.VanKeilegom.03}'s estimates, respectively.}
\end{table}


\section{Applications}~\label{sec.Applications}

\subsection{Dependency of ion-channel recordings: Gramicidin A}~\label{sec:IonChannel}

 Ion channels are proteins regulating the flow of ions across the cell
 membrane by random opening and closing of pores. Typical experiments such 
 as patch-clamp recording would move an electrode close to an ion channel
 allowing that electrical currents flowing through the channel
 can be measured. In this section we consider recordings of 60 secs of 
 gramicidin A; this ion channel trace was recorded at a sampling rate of 
 10kHz and digitized with a low-pass 4-pole Bessel filter at a cut-off frequency 
 of 1kHz, see Figure~\ref{figIntro:a} and \cite{Hotz.etal.13} for further details.

 Common ion channel traces resemble realizations of model \eqref{eq.NPR} 
 where the error model is now the convolution between the (discrete) kernel 
 of the low-pass 4-pole Bessel filter and realizations of i.i.d.~
 normal random variables with zero mean and variance $\sigma^2$.
 That is,
 in this case we can compute the theoretical correlation function of our
 observations. Indeed, we utilize the function \texttt{dfilter}
 from the \texttt{CRAN} package \texttt{stepR} (\cite{Hotz.Sieling.15}),
 and calculate the theoretical correlation between the observations of
 Figure~\ref{figIntro:a}.
 Additionally, we calculate the correlation 
 estimates ($\wh{\rho}_O$, based on \eqref{eq.Delta(121).1}-\eqref{eq.Delta(121).2}), 
 \cite{Herrmann.92}'s ($\wh{\rho}_H$), \cite{Muller.Stadtmuller.88}'s ($\wh{\rho}_M$), 
 adapted-\cite{Rice.84}'s ($\wh{\rho}_R$, set $d=0$ in \eqref{eq.Delta(121).1}) 
 and \cite{Hall.VanKeilegom.03}'s ($\wh{\rho}_{HV}$) and compare them with 
 the theoretical correlation. 
 
\begin{figure}[htb]
 \scalebox{.85}{
  \begin{subfigure}[b]{0.5\linewidth}
    \centering
    \includegraphics[width=\linewidth]{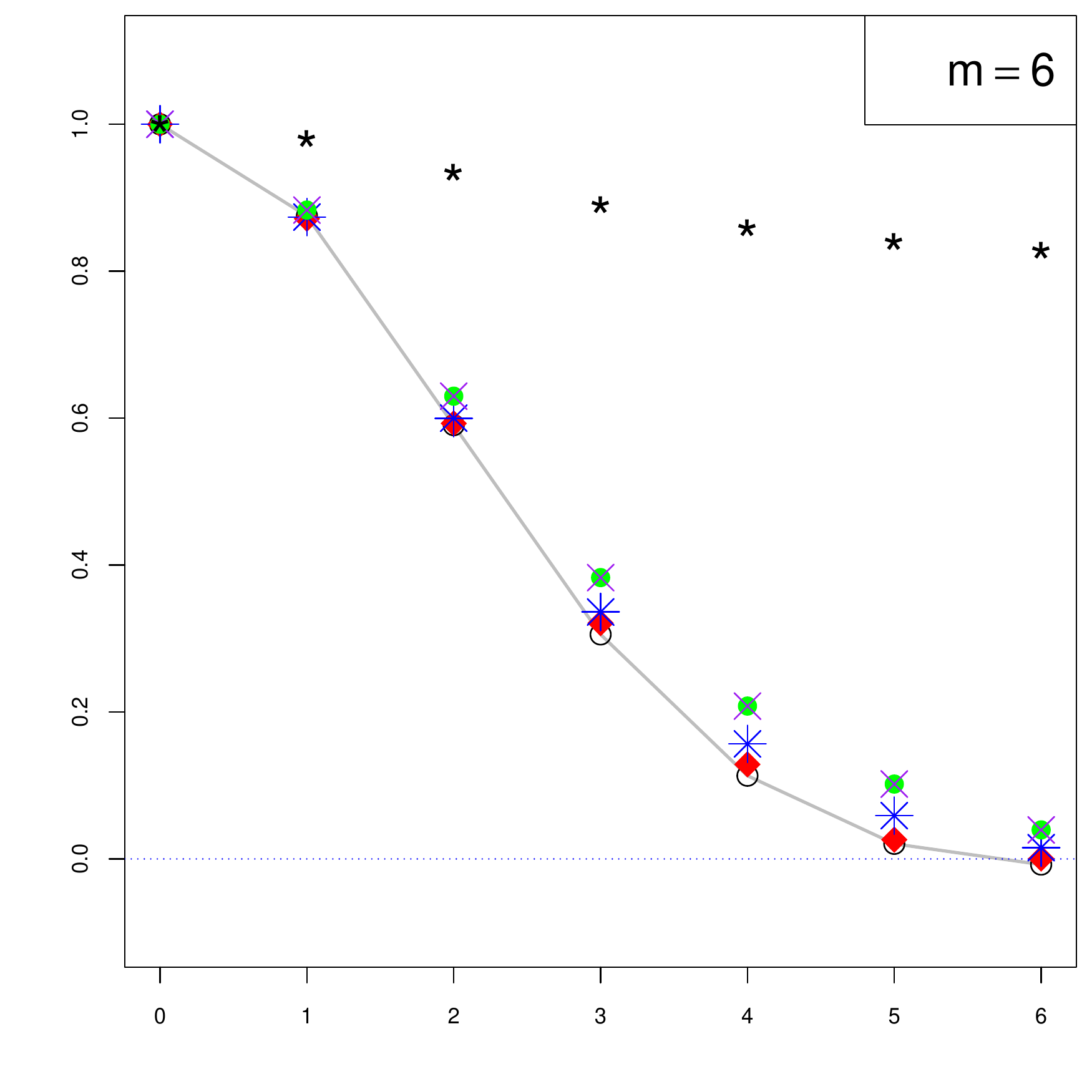} 
    \caption{} 
    \label{fig:ionChannel_gA_acf_a} 
    \vspace{4ex}
  \end{subfigure}
  \begin{subfigure}[b]{0.5\linewidth}
    \centering
    \includegraphics[width=\linewidth]{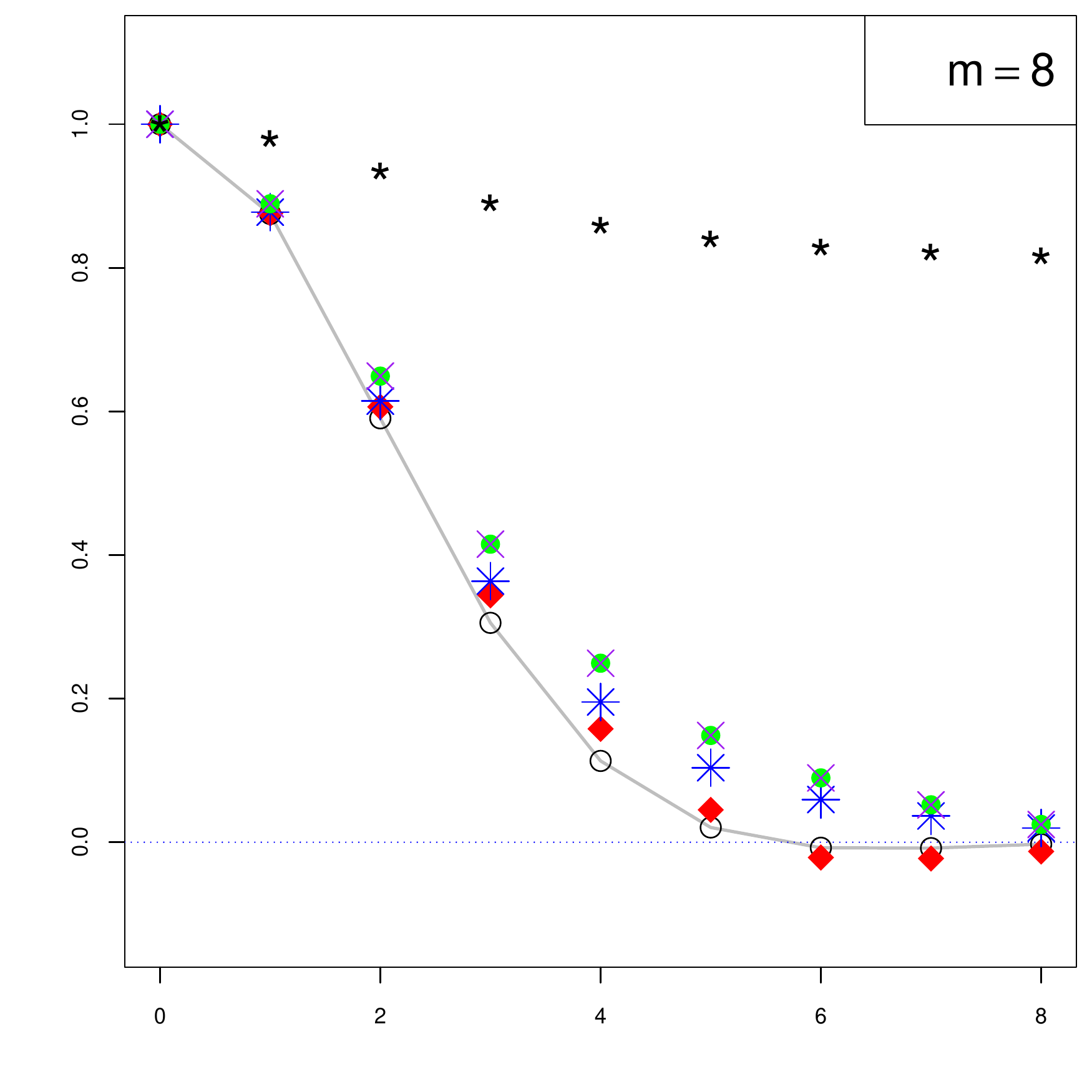} 
    \caption{} 
    \label{fig:ionChannel_gA_acf_b} 
    \vspace{4ex}
  \end{subfigure}
  }\\
  \scalebox{.85}{
  \begin{subfigure}[b]{0.5\linewidth}
    \centering
    \includegraphics[width=\linewidth]{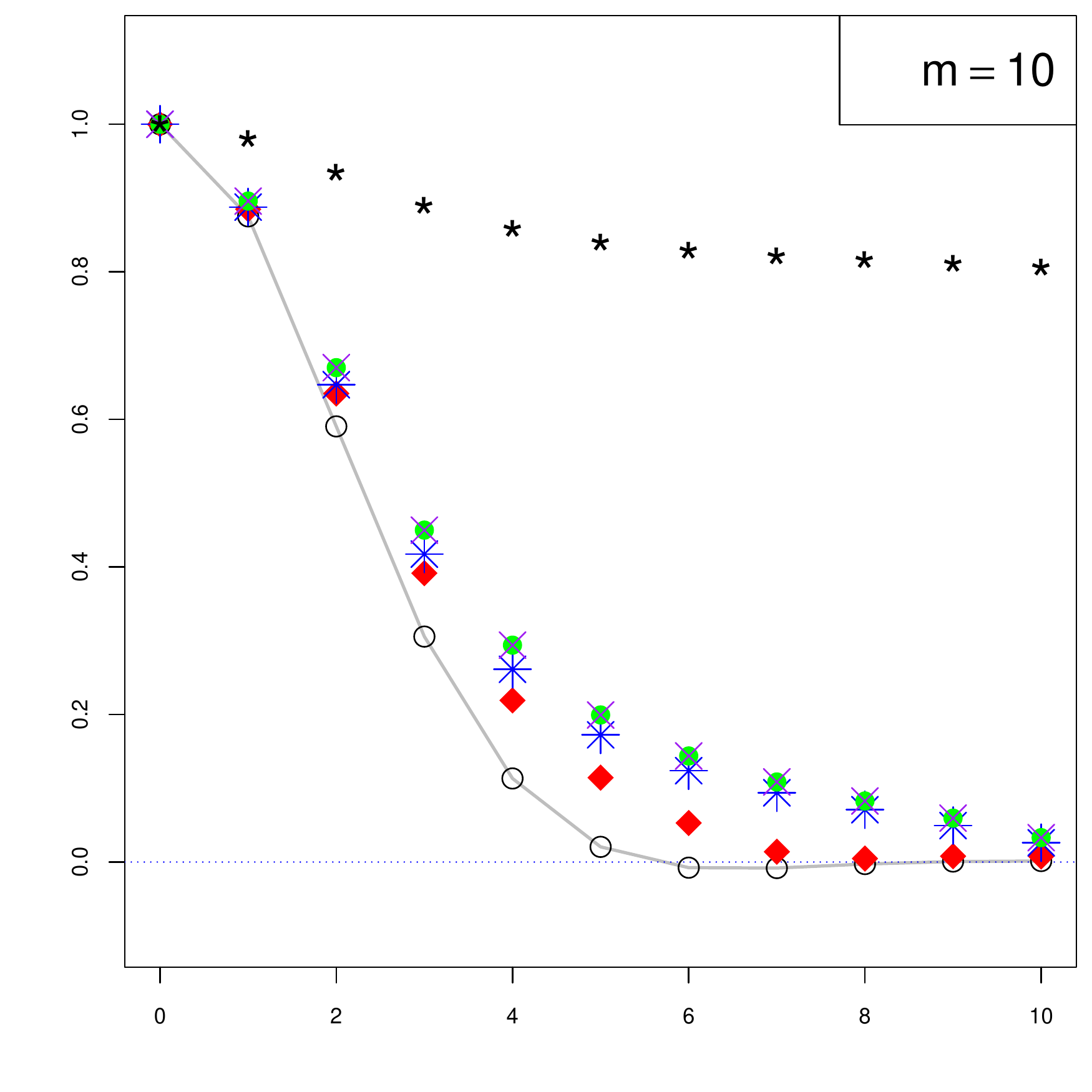} 
    \caption{} 
    \label{fig:ionChannel_gA_acf_c} 
  \end{subfigure}
  \begin{subfigure}[b]{0.5\linewidth}
    \centering
    \includegraphics[width=\linewidth]{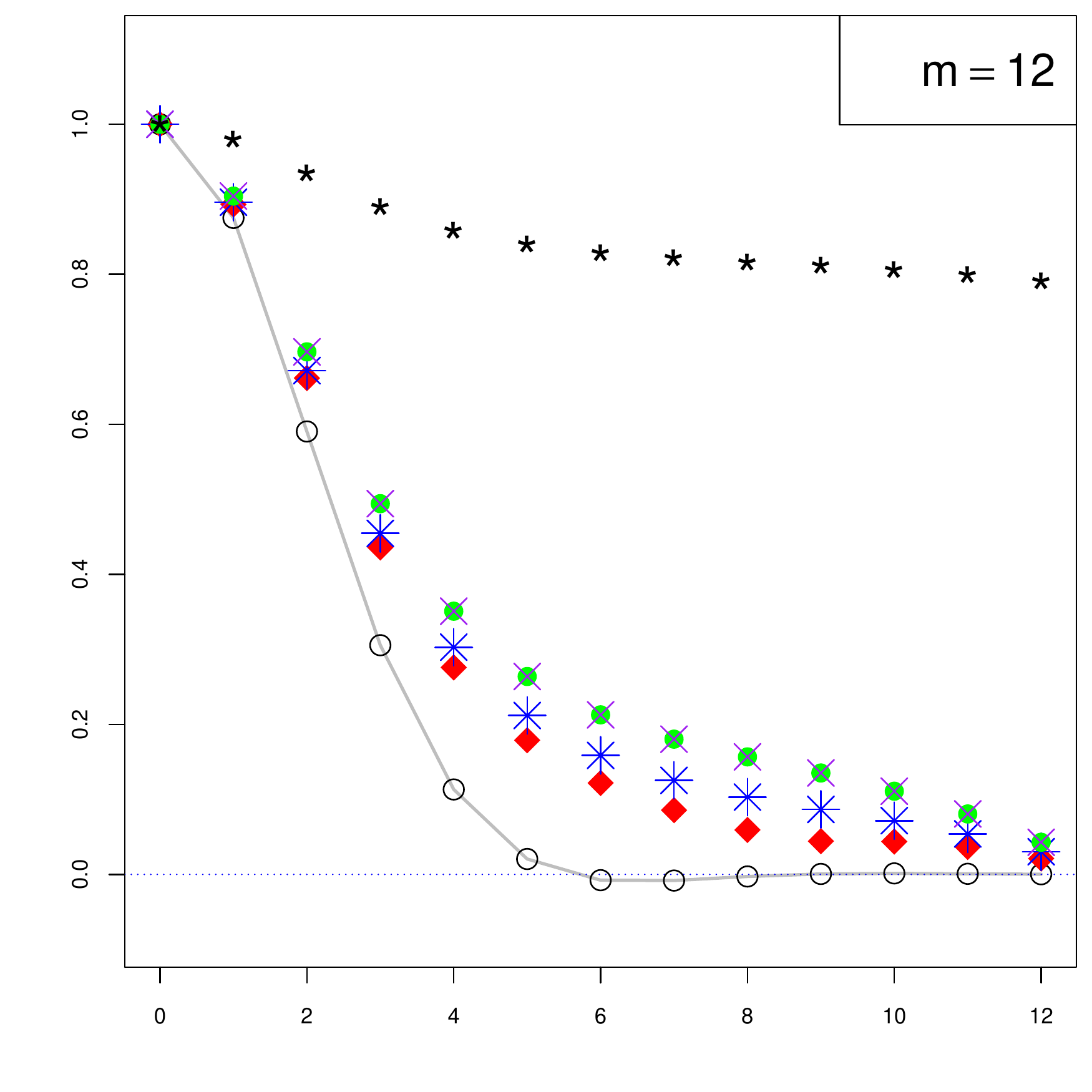} 
    \caption{} 
    \label{fig:ionChannel_gA_acf_d} 
  \end{subfigure} 
  }
  \caption{\footnotesize 
  Correlation estimators from ion channel recordings shown in Figure~\ref{figIntro:a}.
  true ({\large $\circ$}); $\wh{\rho}_O$, optimized difference-based ({\red $\blacklozenge$}); 
  $\wh{\rho}_H$, \cite{Herrmann.92}'s ({\large \blue \textasteriskcentered});
  $\wh{\rho}_M$, \cite{Muller.Stadtmuller.88}'s ({\large \green \textbullet}); 
  $\wh{\rho}_R$, adapted-\cite{Rice.84}'s ({\purple $\times$}); 
  $\wh{\rho}_{HV}$, \cite{Hall.VanKeilegom.03}'s ({\large \black $\star$}).}~\label{fig:ionChannel_gA_acf} 
\end{figure}

 Figure~\ref{fig:ionChannel_gA_acf} displays all these difference-based
 estimates for $m = 6, 8, 10, 12$. Except for $\wh{\rho}_{HV}$ 
 the other estimates show minor quantitative differences between them and to some extent
 they are close to the true correlation. In particular,
 for $m = 6$, the proximity of $\wh{\rho}_O$ with the theoretical correlation
 function is remarkable.
 
 The main goal of this section is to explore an application of our autocovariance 
 estimation techniques in a changepoint problem. To this end, we introduce a method for 
 {\bf JU}mp {\bf S}egmentation of (short-range) {\bf D}epedent data ($\JUSD$).
 This method combines the multiscale changepoint estimator introduced by 
 \cite{Frick.Munk.Sieling.14} with the autocovariance estimates proposed in this paper. 

 \pagebreak
\subsection{Jump segmentation for dependent data}~\label{sec:CPapp}

Now we briefly describe $\JUSD$. Let ${\bf Y}$ denote the vector of observations $y_i = f_i + \varepsilon_i$
following \eqref{eq.NPR} and for simplicity suppose that the errors $\varepsilon_i$
are Gaussian distributed. The multiscale approach to detect a change-point
given in Eqs.~(3)-(4) of \cite{Frick.Munk.Sieling.14} suggests to calculate
all local likelihood ratio test (LRT) statistics $T_i^j({\bf Y}, f_{\mc H})$, $1\leq i < j \leq n$,
and to use the maximum over these statistics as the final test statistic, say $T_n$.
Here $\mc H$ denotes the hypothesis that $f$ takes a fixed value on the
interval $[i/n, j/n]$. Then, estimates and confidence regions for
the change-points of $f$ can be derived from the distribution of $T_n$, which can e.g.~be
obtained via simulations, see \cite{Frick.Munk.Sieling.14} for the
independent case.

In the current situation of $m$-dependent errors, for each $i$ and $j$ the 
local LRT depends also on the \emph{inverse} of the $(j-i+1)\times(j-i+1)$ covariance Toeplitz 
matrix, say $\Sigma_{j-i+1}$.
In the worst case scenario, the dimension of such matrices is $n^2$ and
in order to avoid the expensive calculation of $\Sigma_{j-i+1}^{-1}$ we 
propose a more tractable multiscale statistic. We re-weight
the partial sums of the observations $y_i$ according to the dependency.
More precisely, let $S_i^j = y_{i} + \cdots + y_j$, $i<j$, and consider the modified 
local statistics based on \emph{partial sums}
\[
 {\tilde{T}_i^j({\bf Y}; f_{\mc H})}
 =
 {| S_i^j - f_{\mc H} |^2}/{{\VAR(S_i^j)}},\quad 1\leq i < j\leq n.
\]
Now our proposal ($\JUSD$) consists of utilizing our difference-based estimates $\wh{\gamma}_O$
to estimate ${\VAR(S_i^j)} = (j-i+1)\,\gamma_0 + 2\sum_{k=1}^{m}(j-i+1-k)_+\,\gamma_k$,
here $x_+ = \max(0, x)$.
The corresponding modified version of $T_n$ is denoted by $\tilde{T}_n$.
Under the true error model, Monte Carlo simulations allow us then 
to determine the finite-sample distribution
of $\tilde{T}_n$, and the subsequent estimation of the number of changepoints,
locations and levels of $f$ is carried out as in Eqs.(5)-(6) of \cite{Frick.Munk.Sieling.14}.

For the simulation design of this section we use the signal introduced in 
Section~\ref{sec:ParkErrors_Chakarsignal}
and a $6$-dependent Gaussian process for the error model.
This error model corresponds to the estimated dependence structure of the ion 
channels introduced in Section~\ref{sec:IonChannel}, see Figure~\ref{fig:ionChannel_gA_acf_a}.
To simplify matters in the Monte Carlo experiment to determine the finite
sample distribution of $\tilde{T}_n$, we will use an equivalent MA(6) representation
for the error model. The coefficients of this MA(6) model are obtained via
an application of the Durbin-Levinson algorithm 
(Proposition~5.2.1 of \cite{Brockwell.Davis.06}) to the sequence 
$\wh{\gamma}_h^{(6)}(d_{h,6})$, $h=0,1,\ldots,6$, cf.~\eqref{eq.Delta(121).1}-\eqref{eq.Delta(121).2}.

\begin{figure}[htb]
  \scalebox{.75}{
  \begin{subfigure}[b]{.65\linewidth}
    \centering
    \includegraphics[width=\linewidth]{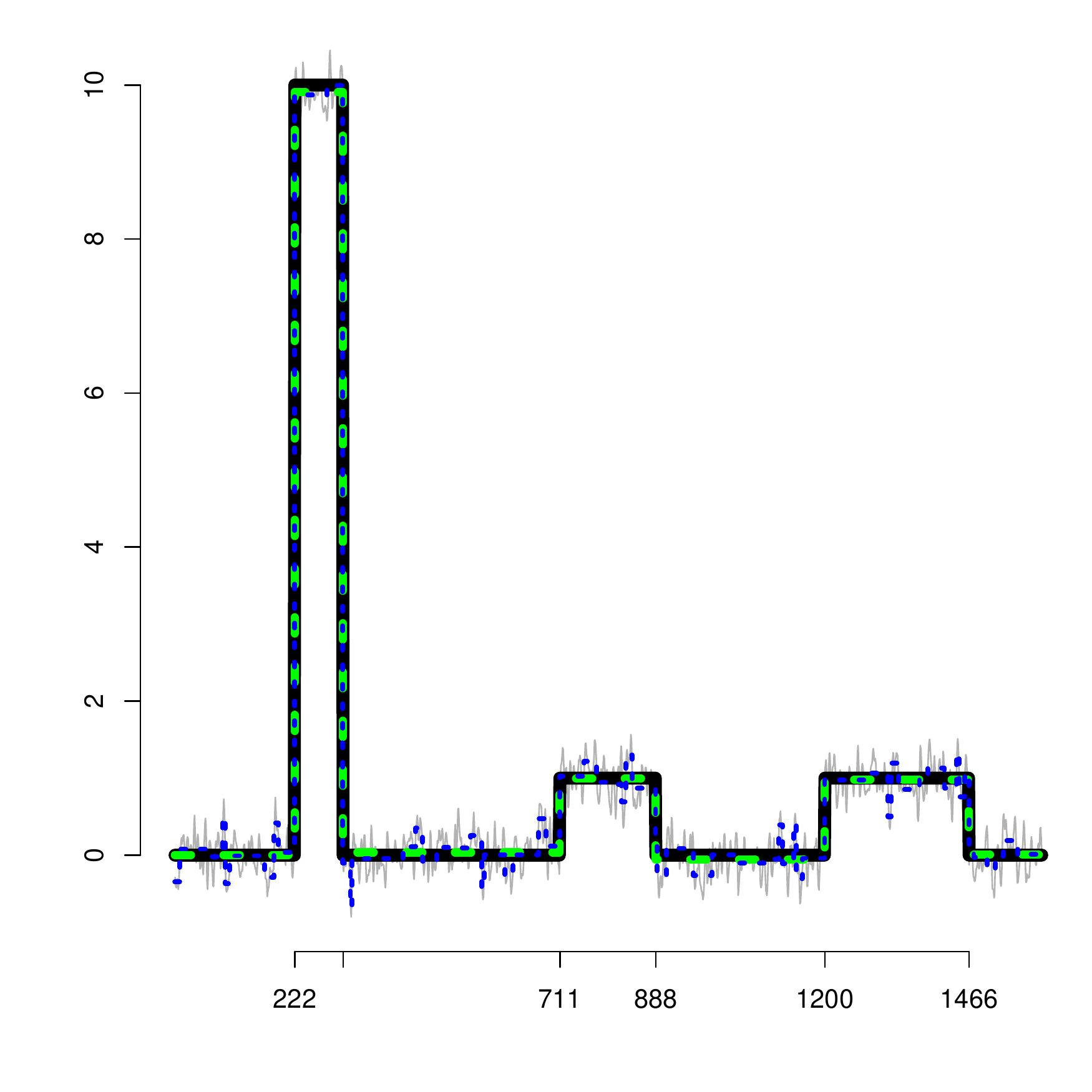} 
    \caption{} 
    \label{fig:changepointApplication_a} 
  \end{subfigure}}\\[.5ex]
  \scalebox{.75}{
  \begin{subfigure}[b]{0.5\linewidth}
    \centering
    \includegraphics[width=.85\linewidth]{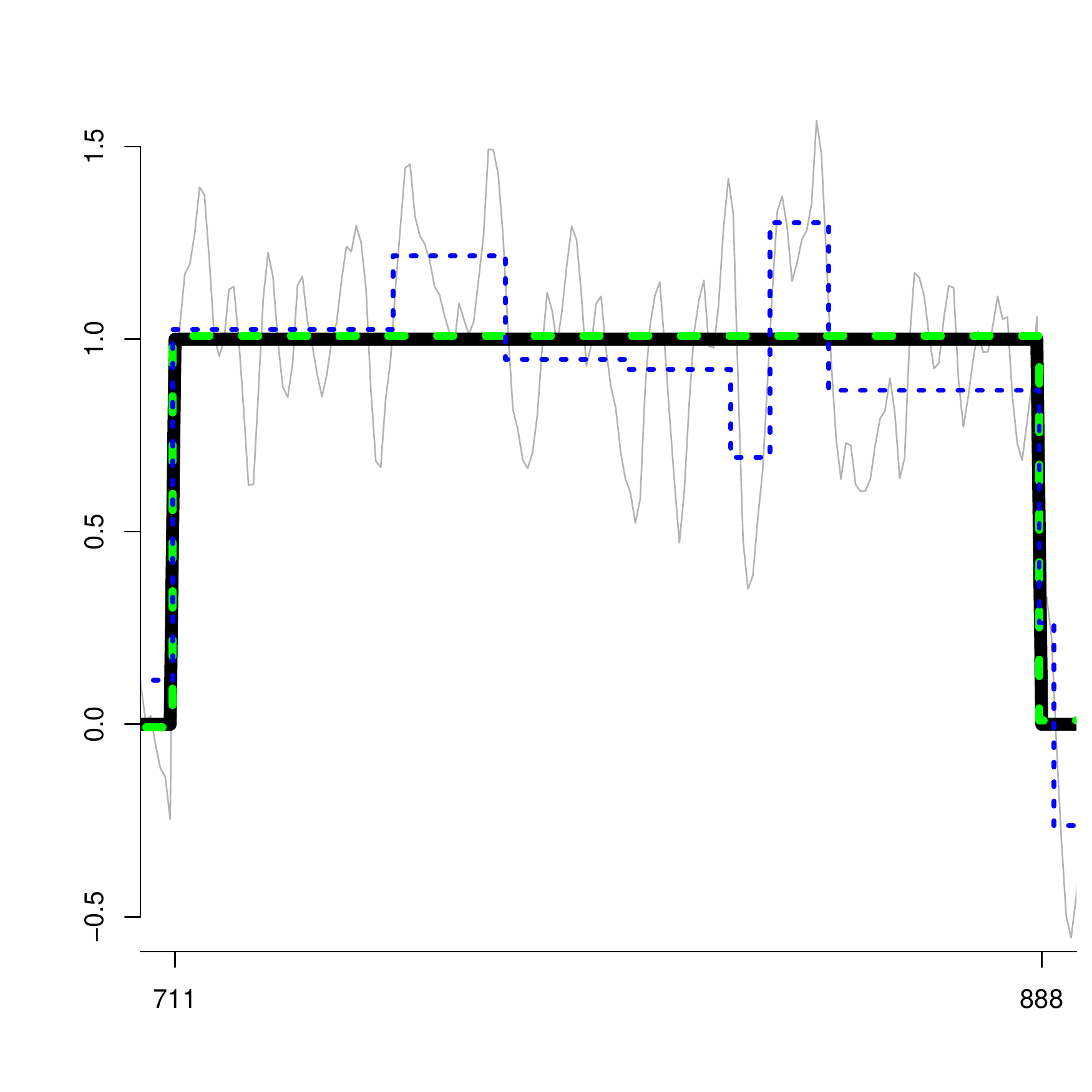} 
    \caption{} 
    \label{fig:changepointApplication_b} 
  \end{subfigure}
  \begin{subfigure}[b]{0.5\linewidth}
    \centering
    \includegraphics[width=.85\linewidth]{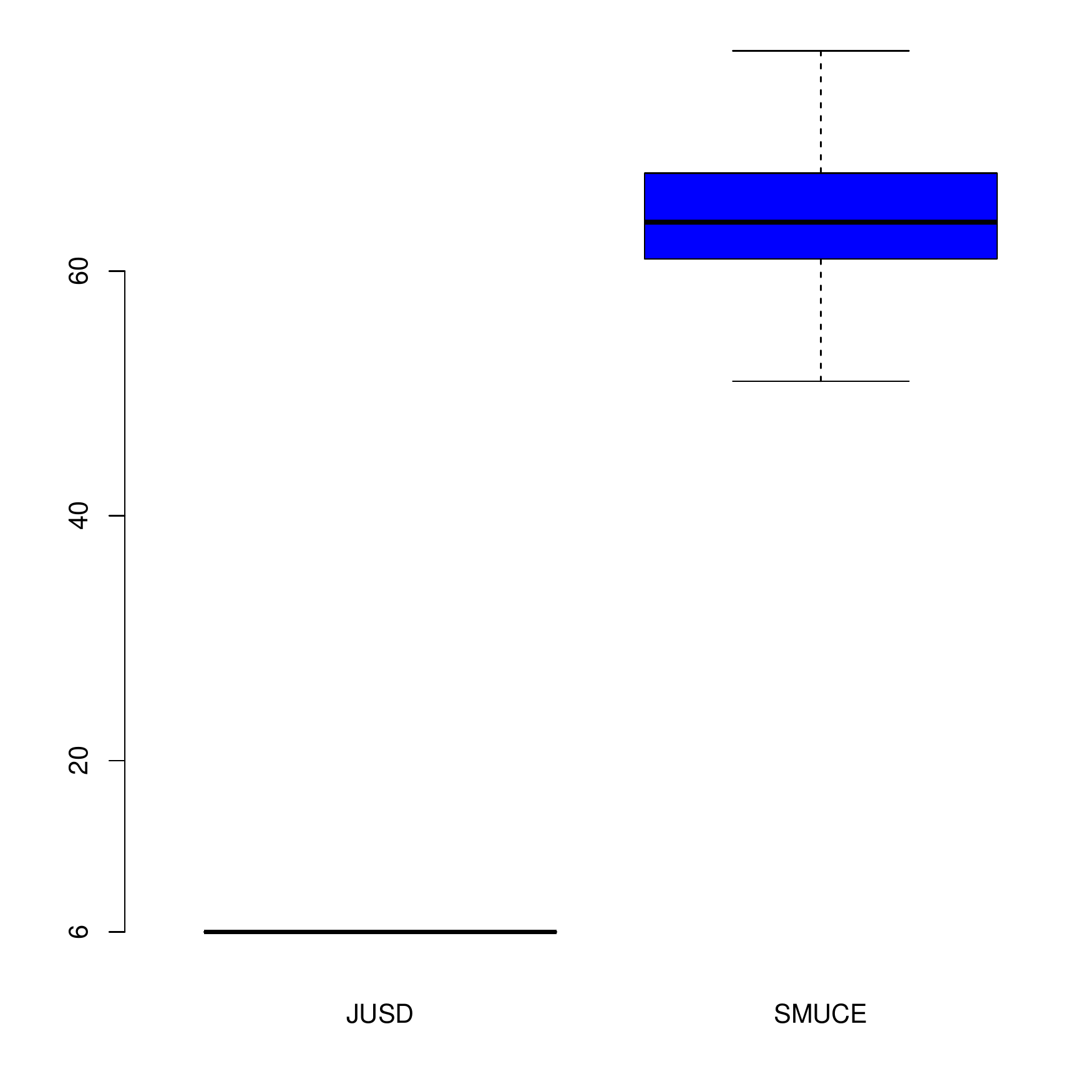} 
    \caption{} 
    \label{fig:changepointApplication_c} 
  \end{subfigure}} 
  \caption{\footnotesize 
  (A) (Gray background) An example of observations $y_i$'s as described
  in Section~\ref{sec:CPapp}.
  ({\black \solidrule}) True piecewise constant signal as described
  in Section~\ref{sec:ParkErrors_Chakarsignal}.
  ({\green \protect\dashedrule})
  {\bf JU}mp {\bf S}egmentation for {\bf D}ependent data as introduced in 
  Section~\ref{sec:CPapp},
  ({\blue \dotsrule})
  {\bf S}imultaneous {\bf MU}ltiscale {\bf C}hangepoint {\bf E}stimator as 
  presented by \cite{Frick.Munk.Sieling.14}.  
  (B) Fit of JUSD and SMUCE on data shown in (A) restricted to the interval 
  $[711, 888)$.
  (C) Boxplot of estimated number of changepoints by $\JUSD$
  and $\SMUCE$; true number of change-points is 6.}
  \label{fig:changepointApplication} 
\end{figure}

This Monte Carlo experiment is performed only once and its output is used to assess
the performance of $\JUSD$ in estimating the signal from 500 samples of size 
$n = 1\,600$. For comparison, we have also utilized the \texttt{R} function \texttt{smuceR} 
which was designed for independent errors by \cite{Hotz.Sieling.15} 
to estimate changepoint locations and levels in the current simulation. 
In each simulation the significance level of the 
estimated number of change points was set to $\alpha = 0.05$ in both methods. 
Figure~\ref{fig:changepointApplication_a} displays the fit
of $\JUSD$ and $\SMUCE$ on one realization of the $y_i$'s
and Figure~\ref{fig:changepointApplication_b} shows the same fits
on a small part of this simulated dataset.
Unlike $\JUSD$, $\SMUCE$ does not take into account the correlation between observations
and this leads to a clear overestimation of the number of changepoints, 
see Figure~\ref{fig:changepointApplication_c}. 

\section*{Acknowledgements}

We are grateful to B.~de Groot at the Max Planck Institute for Biophysical Chemistry at G\"ottingen
and C.~Steinem at the Institute for Organic and Biomolecular Chemistry
at the University of G\"ottingen for providing the datasets analyzed in this paper. 
This paper also benefited from helpful discussions with Timo Aspelmeier,
Florian Pein and Marco Singer at the Institute for Mathematical Stochastics and 
Annika Bartsch at the Institute for Organic and Biomolecular Chemistry
at the University of G\"ottingen. We appreciate the comments of two referees
and the Associate Editor as these improved the contents of this paper. Support of Deutsche 
Forschungsgemeinschaft grants FOR 916-Z and SFB 803-Z2 is gratefully acknowledged.
%
\section*{Supporting information for ``Autocovariance estimation in regression with 
a discontinuous signal and $m$-dependent errors: A difference-based approach''}

\appendix

\section{Proofs and auxiliary results for Section~\ref{sec:Optimization}}~\label{sec.AppendixII}

  Throughout this supplementary materials we use the following
  notation: $n_m := n-2(m+1)$, $n_h := n-h$; $f_i$ denotes $f(x_i)$,
  $v_{i:(i+h)}$ denotes the vector $(v_i \, v_{i+1} \, \cdots \, v_{i+h} )^\top\in\R^{h+1}$
  (we use this notation with $v \in \{y, f, \varepsilon\}$),
  $\inner{\bs{a}}{\bs{b}}$ denotes the inner product between the 
  vectors $\bs{a}$ and $\bs{b}$, ${\bf 1}$ denotes the vector of ones,
  and for $x\in\R^d$, $\|x\|$ denotes its Euclidean norm.\medskip

\subsection{Proofs for Section~\ref{sec.NoGoalTheo}}~\label{sec.ProofsNoGoal}

 We begin with some preliminary results.
 For $l<n$, let ${\bf Q}_1(Y, \bs{w}_l)$ be a difference-based estimator of 
 order $l$ and gap 1, cf.~Eq.~\eqref{eq.DBEk}. With $\wt{D}$ as defined in
 \eqref{eq.Dtilde}
 define the $(l+2)\times(l+2)$ matrix $D:=\wt{D}^\top\,\wt{D}$.
 Observe that the identity
 \[
 \sum_{i=j}^{j+1}\,\left( d_0y_i + d_1y_{i+1} + \cdots + d_l\,y_{i+l} \right)^2
 =
 y_{j:(j+l+1)}^\top\,D\,y_{j:(j+l+1)},\quad j\leq n-2l-3,
 \]
 implies that
\begin{equation}\label{eq.aux.DBE}
 2n_l\,P(\bs{w}_l)\,{\bf Q}_1(Y, \bs{w}_l)
 =
 \inner{\bs{w}_l}{y_{1:(1+l)}}^2
 +
 \sum_{j=1}^{n_l}
 y_{j:(j+l+1)}^\top\,D\,y_{j:(j+l+1)}
 +
 \inner{\bs{w}_l}{y_{(n-l):n}}^2,
\end{equation}
 where $P(\bs{w}_l)= \sum_{i=0}^l d_i^2$. 
 In Eq.~\eqref{eq.aux.DBE} it is not difficult
 to see that $\ME[ \inner{\bs{w}_l}{y_{k:(k+l)}}^2 ]=o(n)$ for $k=1,n-l$ and 
 that for $j<n-l$,
  $\ME[y_{j:(j+l+1)}^\top\,D\,y_{j:(j+l+1)}]=\| \wt{D}f_{j:(j+l+1)} \|^2 + \ME[ \varepsilon_{j:(j+h+1)}^\top\,D\, \varepsilon_{j:(j+h+1)}]$.
  Next, we combine this with Proposition~\ref{lem.trace.DS} and get that,
  \begin{equation}\label{eq.DBE.AsympEV}
  \ME\left[ P(\bs{w}_l) {\bf Q}_1( Y, \bs{w}_l ) \right]
  =
  \bs{w}_l^\top\,\Sigma_{l+1}\,\bs{w}_l 
  + 
  \frac{1}{2n_l}\,\sum_{j=1}^{n_l}\,\| \wt{D}f_{j:(j+l+1)} \|^2 + o(1),
 \end{equation}
 where $\Sigma_{l+1}$ is the $(l+1)\times(l+1)$ autocovariance matrix
 $\Sigma_{l+1} = \left( \gamma_{| i-j |} \right)_{i,j=1,\ldots,l}.$

  \begin{pf}[\textit{\textbf{Proof of Theorem~\ref{theo.NoGoal}}}]
 
  It suffices to consider the difference-based estimator of order $l<n$ and
  gap 1, ${\bf Q}_1(Y, \bs{w}_l)$. 
  For $l\leq m$, due to Lemma~\ref{lem.lim.DF.PC}, 
  \eqref{eq.DBE.AsympEV} becomes
  \begin{equation}\label{eq.DBE.AsympEV2}
    \ME\left[ P(\bs{w}_l) {\bf Q}_1( Y, \bs{w}_l ) \right]
    =
    B_l + \mc O \left( n^{-1}\,J_K\, \sum_{k=1}^l\left( \sum_{j = k}^l\, d_j \right)^2\right),
  \end{equation}
  where
  \begin{equation}~\label{eq.expValue_DBE}
  B_l
  = 
  \bs{w}_l^\top\,\Sigma_{l+1}\,\bs{w}_l
  =
  \gamma_0(d_0^2 + \cdots + d_l^2) + 2\gamma_1(d_0d_1 + \cdots + d_{l-1}d_l) + \cdots + 2\gamma_ld_0d_l.  
  \end{equation}

  From now on in this proof we assume that $J_K = o(n)$ and disregard
  the second summand in the right-hand side of \eqref{eq.DBE.AsympEV2}.
  Since constraint $\inner{\bs{w}_l}{\bs{1}}=0$ implies that at least 
  one of the weights $d_j$ is nonzero, for simplicity
  from now on we assume that $d_0\neq 0$.
  According to Lemma~\ref{lem.noConsistency_h_less_m} in this case 
  there does not exist any constant $c$ such that $B_l = c\,\gamma_0$, that is,  
  no difference-based estimate of the form ${\bf Q}_1(Y, \bs{w}_l)$ can 
  be an asymptotically unbiased estimate for $\gamma_0$ for $l\leq m$.
      
 Next, suppose that $m < l \leq n$.
 In what follows, for simplicity, let us assume that $n=N(g+1)$ for some integer $N\geq 1$
 and $g=m$.
 Observe that in this case Proposition~\ref{lem.trace.DS} and \eqref{eq.DBE.AsympEV2}
 yield, $\ME\left[ p(\bs{w}_l) {\bf Q}_1( Y, \bs{w}_l ) \right] = B_l + o(1)$. 
 Due to $m$-dependency, the covariance matrix $\Sigma_{l+1}$
 appearing in $B_l$ is an $(m+1)$-banded Toeplitz matrix, 
 i.e., the $(i,j)$ entry of $\Sigma_{l+1}$ 
 is given by $\gamma_{|j-i|}\neq 0$ if $|j-i|\leq m$, and
 outside the $(m+1)$-diagonal the entries of $\Sigma_{l+1}$ are equal to 0.
 Suppose that $\bs{w}_l$ has entries $d_0=1$ and 
 $d_j\neq0$, for $j=k(g+1)$ with $k=1,\ldots,N-1$, and $d_j=0$ otherwise. 
 Clearly, $B_l=P(\bs{w}_l)\,\gamma_0$. 
 Now we show that any asymptotically unbiased estimate for $\gamma_0$
 is necessarily of the form just described. Indeed, it suffices to consider 
 the vector $\bs{w}_l^\ast$, whose entries are identical to those of $\bs{w}_l$
 except for $d_\kappa\neq 0$ for some $\kappa\in\{1,\ldots,m\}$.
 In this case, due to the form of the covariance matrix $\Sigma_{l+1}$, 
 $(\bs{w}_l^\ast)^\top\,\Sigma_{l+1}\,\bs{w}_l^\ast = c_1\gamma_0 + 2\,d_\kappa\,\gamma_\kappa$,
 for some constant $c_1$. Since $\gamma_\kappa\neq 0$, no difference-based estimate
 of the form ${\bf Q}_1(Y, \bs{w}_l^\ast)$ can be an asymptotically 
 unbiased estimate for $\gamma_0$. 
 
 Note that the arguments above hold also for $g\geq m$.
 Thus, we have shown that for ${\bf Q}_1(Y, \bs{w}_l)$ to be
 an (asymptotically unbiased) estimate for $\gamma_0$,
 the vector of weights $\bs{w}_l$ must have the form
 $\bs{w}_l = ( \bs{v}_0 \quad \bs{v}_1 \quad \cdots \quad \bs{v}_{N-1} )^\top$,
  where $\bs{v}_i = ( d_{i\cdot g} \quad 0 \quad \cdots \quad 0)^\top\in\R^{g+1}$,
  $i=0,\ldots,N-1$; $d_0\neq 0$, $d_{k\cdot g}\neq 0$ for some $1\leq k\leq N-1$, and 
  $\sum_{i=0}^{N-1}d_{i\cdot g}=0$. 
  This completes the proof.
 \end{pf}
\medskip

\begin{lem}~\label{lem.noConsistency_h_less_m}
  Suppose that the conditions of Theorem~\ref{theo.NoGoal} hold.
  Let ${\bf Q}_1(Y, \bs{w}_l)$ be the difference-based estimator of order $l<n$ and
  gap 1, cf.~\eqref{eq.DBEk}.
  Then for 
  $l\leq m$, there does not exist any constant $c\neq 0$ such that 
  $B_l = \bs{w}_l^\top\,\Sigma_{l+1}\,\bs{w}_l = c\,\gamma_0$, where $\Sigma_{l+1}$
  is defined by \eqref{eq.DBE.AsympEV}.
\end{lem}
  \begin{pf}
   We use induction over $l$. For $l=1$, $B_1 = \gamma_0(d_0^2 + d_1^2) + 2\gamma_1d_0d_1$,
   cf.~\eqref{eq.expValue_DBE}.
   Since by assumption $\gamma_1\neq 0$ (and $d_0\neq 0$), $d_1$ is necessarily
   equal to zero but this implies that $\gamma_0=0$ to fulfill \eqref{eq.a}.
   This contradiction shows that the claim holds for $h=1$. Then we assume 
   that the claim holds for $l = k \leq m-1$. Let $\tilde{k} = k + 1$ and 
   now note that for $B_{\tilde{k}} = c\,\gamma_0$ to hold for some $c\neq 0$, 
   necessarily $\gamma_{\tilde{k}}d_0d_{\tilde{k}}=0$. Since
   $\gamma_{\tilde{k}}\neq 0$ and $d_0\neq 0$, necessarily $d_{\tilde{k}}=0$. 
   The latter shows that $B_{\tilde{k}} = B_k$. Hence, the claim holds for 
   $l = k+1$ and this completes the proof.
  \end{pf}

\begin{pf}[Proof of Lemma~\ref{lem.lim.DF.PC}]
  For $j < n-l$ define $s_{j,l} = \inner{\bs{w}_l}{f_{j:(j+l)}}^2$ and note that
  $\| \wt{D}f_{j:(j+l+1)}^\top \|^2 = s_{j,l} + s_{j+1,l}$, see \eqref{eq.Dtilde}.
  In what follows we only consider $s_{j,l}$ since $s_{j+1,l}$ can be handled
  similarly.
  
  Under the convention $t_k = \lfloor n \tau_k \rfloor$, $f(x_i) = a_k$ if and only
  if $t_{k-1}\leq i < t_k$. Then
 \begin{equation}\label{eq.sum.sjh}
  \sum_{j=1}^{n_l}s_{j,l}
  \leq
  \sum_{i=0}^{K-1}
  \left[ \sum_{j = t_i}^{t_{i+1}-l-1}s_{j,l}  
  + 
  \sum_{j = t_{i+1}-l}^{t_{i+1}-1}s_{j,l}\right].
 \end{equation}
 
  Then note that for any $i\in\{0, \ldots, K-1\}$,
 \[
  \sum_{j = t_i}^{t_{i+1}-l-1}s_{j,l} 
  =
  a_i^2\inner{\bs{w}_l}{\bs{1}}^2(t_{i+1}-t_i-l-1)
 \]
 and utilizing that $d_0+\cdots+d_l=0$,
 \[
  \sum_{j = t_{i+1}-l}^{t_{i+1}-1}s_{j,l}
  =
  \sum_{j = t_{i+1}-l}^{t_{i+1}-1}( d_0f_j + \cdots + d_lf_{j+l} )^2\\
  =
  (a_i-a_{i+1})^2\,\sum_{k=1}^l\left( \sum_{j = k}^l\, d_j \right)^2.
 \]
 Substituting these expressions into \eqref{eq.sum.sjh} we obtain the
 right-hand side of \eqref{eq.lem.BIAS.DBE}. This completes the proof.
\end{pf}

 \begin{prop}\label{lem.trace.DS}
  For $l< n$, let $\bs{w}_l\in \R^{l+1}$ be the vector of weights in 
  ${\bf Q}_1(Y, \bs{w}_l)$, cf.~\eqref{eq.DBEk}.
  Let $D=\wt{D}^\top\,\wt{D}$ where $\wt{D}$ is defined by \eqref{eq.Dtilde}.
  Then
  \[ 
  \ME[ \varepsilon_{j:(j+l+1)}^\top\,D\, \varepsilon_{j:(j+l+1)}]
  =
  2\,\bs{w}^\top_l\,\Sigma_{l+1}\,\bs{w}_l.
  \]
 \end{prop}
 \begin{pf}
  Let $\Sigma_{l+1}$ be the matrix defined in \eqref{eq.DBE.AsympEV}
  and note that
  \[
  \ME[ \varepsilon_{j:(j+l+1)}^\top\,D\, \varepsilon_{j:(j+l+1)}]
  =
  \mbox{tr}\{ D \Sigma_{l+2} \} 
  = 
  \mbox{tr}\{ \wt{D}\,\Sigma_{l+2}\,\wt{D}^{\top} \}.
  \]
  Then, observe that the $2\times 2$ matrix $\wt{D}\,\Sigma_{l+2}\,\wt{D}^{\top}$ can be 
  written as:
  \begin{align*}
  \wt{D}\,\Sigma_{l+2}\,\wt{D}^{\top}
  &=
  \begin{pmatrix}
    \bs{w}^\top_l\,\Sigma_{l+1} & \inner{ \bs{w}_l }{ \gamma_{(l+1):1} }\\
    \inner{ \bs{w}_l }{ \gamma_{(l+1):1} } & \bs{w}^\top_l\,\Sigma_{l+1}
  \end{pmatrix}\,
  \begin{pmatrix}
    \bs{w}_l & 0\\
    0 & \bs{w}_l
  \end{pmatrix},
  \end{align*}
  where 
  $\gamma_{(l+1):1} := ( \gamma_{l+1} \quad \gamma_{l} \quad \cdots \quad \gamma_2 \quad \gamma_1 )^\top\in\R^{l+1}$.	
  A straightforward calculation yields, 
  $\mbox{tr}\{ \wt{D}\,\Sigma_{l+2}\,\wt{D}^{\top} \} = 2\bs{w}^\top_l\,\Sigma_{l+1}\,\bs{w}_l$.
 \end{pf}\smallskip

\subsection{Proofs for Section~\ref{sec.bias.minimizer}}~\label{sec.Proofs.bias.minimizer}

  We will assume that the conditions of Theorem~\ref{theo.MSE} hold.
  Also we will use the following notation:
  $\chi_h(f_i) := f_i-f_{i+h}$, for $d\in \R$,
  $\delta_i(d):= s_0(i) + d\,s_{2(m+1)}(i)$ where for 
  $k\geq0$, $s_k(i) := f_{i+k}-f_{i+m+1}$, 
  $\eta_i(d):=\varepsilon_i-\varepsilon_{i+m+1}+d( \varepsilon_{i+2(m+1)} - \varepsilon_{i+m+1})$,
  set $P(d) = 2(d^2+d+1)$ and for given $\tau_i,\tau_j \in [0,1]$, $1\leq L,R \leq n$,
  \begin{equation}\label{eq.Intervals}
   I_{\tau_i, \tau_j}^{L,R} := [\tau_i-L/n,\tau_j-R/n).
  \end{equation}
  Also, $I_{\tau_i}^{L,R} := I_{\tau_i, \tau_i}^{L,R}$.
  Under the convention $t_j = \lfloor n\tau_j \rfloor$, we will denote
  $I_{t_i, t_j}^{L,R} := [t_i-L,t_j-R)$ and use that $i/n \in I_{\tau_i, \tau_j}^{L,R}$
  if and only if $i\in I_{t_i, t_j}^{L,R}$ without further mention.
  
 The following identities are of great use in what follows: for any integers 
 $r,s,u$ and $v$,
 \begin{align}
	\ME[ \varepsilon_{r}^2\,\varepsilon_s^2 ] 
	&= 
	\gamma_0^2 + 2\,\gamma_{|r-s|}^2\label{eq.ExpValE2E2}\\
	\ME[ \varepsilon_{r}^2\,\varepsilon_u\,\varepsilon_v ] 
	&=
	\gamma_0\gamma_{|u-v|} + 2\gamma_{|r-u|}\gamma_{|r-v|} \label{eq.ExpValE2EE} \\
	\ME[ \varepsilon_{r}\,\varepsilon_s\,\varepsilon_u\,\varepsilon_v ] 
	&=
	\gamma_{|r-s|}\gamma_{|u-v|} + \gamma_{|r-u|}\gamma_{|s-v|} + \gamma_{|r-v|}\gamma_{|s-u|},\label{eq.ExpValEEEE}
 \end{align}
 cf.~Theorem~3.1 of \cite{Trianta.03}.
 \begin{lem}\label{lem.ExpValEtai2}
  Let $i\geq 1$, $l\geq 1$, $0\leq h\leq (m+1)$ and define 
  $E_{i,l,h}=d_0\,\varepsilon_i+d_1\varepsilon_{i+h} + d_2\varepsilon_{i+2h}
  +\cdots+d_l\varepsilon_{i+lh}$. Then,
  \[
   \ME[E_{i,l,h}^2] 
   =
   \gamma_0(d_0^2+d_1^2+d_2^2+\cdots+d_l^2) + 2\,\sum_{j=0}^{l-1}\sum_{k=j+1}^{l}\,d_j\,d_k\,\gamma_{|j-k|\,h}
  \]
 \end{lem}
 \begin{pf}
  Write $E_{i,l,h}^2= A_{i,l,h} + 2 B_{i,l,h}$, where
	\begin{equation}\label{eq.Etai2Aux}
	 A_{i,l,h}
	 =
	 \sum_{j=0}^l\,d_j^2\varepsilon_{i+jh}^2,\quad
	 B_{i,l,h}
	 =
	 \sum_{j=0}^{l-1}\,x_j(i,l,h),\quad
	 x_j(i,l,h)
	 =
	 d_j\varepsilon_{i+jh}\,\,\sum_{k=j+1}^l\,d_k\,\varepsilon_{i+kh}.	 
	\end{equation}
	The result follows by noticing that due to stationarity,
	for any integers $i$, $j$, $k$ and $h$, $\ME[\varepsilon_{i+jh}^2]=\gamma_0$
	and $\ME[ \varepsilon_{i+jh} \varepsilon_{i+kh}]=\gamma_{|j-k|h}$.
  \end{pf}

\begin{cor}\label{cor.ExpEtai2}
 For $l=2$, $d_0=1$, $d_1=-(d+1)$ and $d_2=d$, $\ME[\eta_i^2(d)] = 2(d^2+d+1)\,\gamma_0$.
\end{cor}

\begin{lem}\label{lem.ExpValEtai4}
  For $i\geq 1$, $l\geq 1$,  
  $\ME[ E^4_{i,l,m+1} ] = 3\,\gamma_0^2\,(d_0^2+d_1^2+\cdots+d_l^2)^2$.
\end{lem}
\begin{pf}
 From \eqref{eq.Etai2Aux}, $E^4_{i,l,m+1} = A^2_{i,l,m+1} + 4A_{i,l,m+1}B_{i,l,m+1} 
 + 4B^2_{i,l,m+1}$.
 Note that \eqref{eq.ExpValE2EE} implies that $\ME [A_{i,l,m+1}\,B_{i,l,m+1}] = 0$.
 That is, 
 \begin{equation}\label{eq.ExpValEtai4_a}
  \ME[ E^4_{i,l,m+1} ]
  =
  \ME[ A^2_{i,l,m+1} ] + 4 \ME[ B^2_{i,l,m+1} ].
 \end{equation}
  Eq.~\eqref{eq.ExpValE2E2} yields that for any integers
  $i$, $j$ and $k$, $\ME [\varepsilon_{i+j(m+1)}]=3\gamma_0^2$
  and $\ME[ \varepsilon_{i+j(m+1)}\,\varepsilon_{i+k(m+1)} ]=\gamma_0^2$. Consequently,
  \begin{equation}~\label{eq.ExpValEtai4_b}
   \ME[A^2_{i,l,m+1}] 
   =
   \gamma_0^2\left( 3\sum_{j=0}^l\,d_j^4 + 2 \sum_{k,j}\,d_k^2\,d_j^2 \right).
  \end{equation}

 Next, Eq.~\eqref{eq.ExpValEEEE} implies that
 for $r\neq s$, $\ME [ x_r(i,l,m+1) x_s(i,l,m+1) ] = 0$, see
 \eqref{eq.Etai2Aux} for definition of $x_{(\cdot)}(i,l,m+1)$.
 Hence,
 \begin{equation}~\label{eq.ExpValEtai4_c}
  \ME [ B^2_{i,l,m+1} ]
  =
  \sum_{j=0}^l\,\ME [ x_j^2(i,l,m+1) ]
  =
  \gamma_0^2\sum_{k,j}\,d_k^2\,d_j^2.
 \end{equation}
  The last equality follows from \eqref{eq.ExpValE2E2}. The result follows
  by plugging \ref{eq.ExpValEtai4_b} and \ref{eq.ExpValEtai4_c} into \ref{eq.ExpValEtai4_a}.
\end{pf}


 \begin{lem}\label{lem.A}
 Suppose that the conditions of Theorem~\ref{theo.MSE} hold.
 Then,
	\begin{align}
		\frac{\bf A}{8}
		&:=
		\sum_{i=1}^{n_m-1}\sum_{j=i+1}^{n_m}\delta_i(d)\delta_j(d)\ME[ \eta_i(d) \eta_j(d) ]
		=
		\left[-d(d+1)^2(m+1) + \sum_{h=1}^m\,q_h^{(m)}(d)\rho_h\right] \gamma_0\,J_K,
		~\label{eq.A8}
	\end{align}
	where $q_h^{(m)}(d) = [2(m+1) - 3h](d^4 + 1) + d^2 h$ and
	$\rho_h = \gamma_h/\gamma_0$.
 \end{lem}
 \begin{pf}
  For $d\in\R$, let $S_r(d) = \sum_{i=1}^{n_m-r}\,\delta_i(d)\delta_{i+r}(d)$, $1\leq r< n_m$.
  First observe that due to stationarity,
  for any $i\geq 1$ and $h\geq 1$, 
  $\Psi(h,d):=\ME[\eta_i(d)\,\eta_{i+h}(d)]=2\,(d^2+d+1)\gamma_h-(d+1)^2\gamma_{|h-(m+1)|}+d\gamma_{|h-2(m+1)|}$.
  Then due to $m$-dependency, $\Psi(h,d)=0$ for all $h>3m+2$. Consequently,
  we can write
\begin{equation}\label{eq.A8Aux}
	\frac{\bf A}{8}
	=
	\sum_{r=1}^{3m+2}\,\Psi(r)\,S_r.
\end{equation}

	Next, we present the details on how to compute $S_0(d)$.
%
%
	 Utilizing \eqref{eq.DistBetweenJumps} and \eqref{eq.Intervals}, we can 
	 show that for given $\tau_j$,
 	 \[
	  \sum_{x_i\in I_{\tau_j}^{2(m+1),m+1}}\,\delta_i^2(d) = (m+1)d^2(a_{j+1} - a_j)^2,\quad
	  \sum_{x_i\in I_{\tau_j}^{m+1,0}}\,\delta_i^2(d) = (m+1)(a_{j+1} - a_j)^2.
	 \]
	Observe that 
	$S_0(d) 
	= 
	\sum_{j=0}^{K-1}\,\sum_{x_i\in I_{\tau_j}^{2(m+1),m+1} \cup I_{\tau_j}^{m+1,0} }\,\delta_i^2(d)
	=
	(m+1)(d^2+1)\,J_K$.
	The key part in obtaining the summation $S_r(d)$, $r\geq 1$, consists of splitting it
	as shown above (and using \eqref{eq.DistBetweenJumps}-\eqref{eq.Intervals}). 
	Thus, we can show that
	 $S_r(d) = T_r(d)\,J_K$ where
  \begin{equation}\label{eq.FormulaSr}
   T_r(d)
   =   
   \begin{cases}
    (m+1-r)d^2 + rd + (m+1-r) & \mbox{ for }r=0,\ldots,m\\
    d\,(2(m+1)-r) & \mbox{ for }r=m+1,\ldots,2m+1\\
    0 & \mbox{ for }r\geq 2(m+1)
   \end{cases}.
  \end{equation}
  In order to get \eqref{eq.A8}, substitute \eqref{eq.FormulaSr} into \eqref{eq.A8Aux} 
  and arrange terms. This completes the proof.
 \end{pf}
	
 \begin{lem}\label{lem.B}
 Suppose that the conditions of Theorem~\ref{theo.MSE} hold. Then
  \begin{align}
  {\bf B} 
  &:= 
  2\sum_{i=1}^{n_m-1}\sum_{j=i+1}^{n_m}\ME[\eta_i^2(d)\,\eta_j^2(d)]
  =
  8n_m^2\,(d^2+d+1)^2\gamma_0^2 + 2\,\sum_{r=1}^{3m+2}(n_m-r)\,\Lambda_r(d)\geq 0.\label{eq.B}
 \end{align}
 where
 \begin{align}
  \Lambda_r(d;\gamma_{(\cdot)})
  &=
  8\,(d^2+d+1)^2\gamma_h^2 + 2(1+d)^4\gamma_{| h-(m+1) |}^2 + 2d^2\gamma_{| h-2(m+1) |}^2\notag\\
  &-
  4(1+d)[ (1+d)^3 + (d^3+d^2+d+1) ]\gamma_h\,\gamma_{|h-(m+1)|}\notag\\
  &-
  4d(d+1)^2\gamma_{|h-(m+1)|}\,\gamma_{|h-2(m+1)|}.\label{eq.H}
 \end{align}
 \end{lem}
 \begin{pf}
  Straightforward calculations and Eqs.~\eqref{eq.ExpValE2E2}, \eqref{eq.ExpValE2EE} 
  and \eqref{eq.ExpValEEEE} yield that for $i\geq 1$ and $r\geq 0$,
  \[
   \ME[\eta_i^2(d)\,\eta_{i+r}^2(d)] = 4\,(d^2+d+1)^2\gamma_0^2 + \Lambda_r(d).
  \]
  Arrange terms and use $m$-dependency to get
  \[
   \frac{\bf B}{2}=\sum_{r=1}^{n_m-1}\,\sum_{s=1}^{n_m-r}\ME[\eta_s^2(d)\,\eta_{s+r}^2(d)].
  \]
  The result now follows by noticing that $\Lambda_r(d)=0$ for all $r\geq 3(m+1)$. 
 \end{pf}

 \begin{lem}\label{lem.MaxCorrProcesses}
  Suppose that the assumptions of Theorem~\ref{theo.MSE} hold.
  Additionally, assume that the correlation function $\rho_h=\gamma_h/\gamma_0$ 
  satisfies that $\rho_h=\rho\in ( \max\{\,-1,-8/(3m^2)\,\},1)$, $1\leq h\leq m$.
  Then $p_1(d;\gamma_{(\cdot)})$ and $\BIAS^\ast[\wh{\gamma}_0^{(m)}(d)]$
  are minimized at $d=1$.
 \end{lem}
  \begin{pf}
  Since $\BIAS[\wh{\gamma}_0^{(m)}(d)]$ is minimized at $d=1$, 
  cf.~Theorem~\ref{cor.BIASgamma0}, we only need
  to focus on minimizing $p_1(\cdot; \gamma_{(\cdot)})$.
  Let $Q(d)=d^2+d+1$.
  It is easily seen that
  \[
    p_1(d;\gamma_{(\cdot)})
    =
    \frac{m+1}{Q^2(d)}\left[ 2(d^4+1) + m(d^4+d^2+1)\sum_{h=1}^{m}\rho_h \right].
  \]
  For $\rho=0$ (indepent observations) the result follows since 
  $\argmin_{d\in \R}(d^4+1)/Q^2(d) = 1$. For $\rho > 0$,
  we have that $\argmin_{d\in \R}(d^4+d^2+1)/Q^2(d) = 1$
  and hence for $d\in\R$, $p_1(d;\gamma_{(\cdot)}) \geq p_1(1;\gamma_{(\cdot)})$.
%
  For $\rho < 0$, note that
  \[
  \frac{\partial}{\partial d}\,p_1(d;\gamma_{(\cdot)})
  =\frac{2(m+1)(d^2-1)( d^2(\rho\,m^2+2) + d(\rho\,m^2+4) + \rho\,m^2+2 )}{Q(d)^3}, 
  \]
  It is immediate that on $\R$, the critical points of $p_1$ are -1 and
  1. For $\rho \in (-8/(3m^2),0)$, $\frac{\partial^2}{\partial d^ 2}\,p_1(-1;\gamma_{(\cdot)}) = -4\rho\,m^2 >0$
  and $\frac{\partial^2}{\partial d^ 2}\,p_1(d;\gamma_{(\cdot)})=4(9 \rho\,m^2 + 24)/81>0$,
  i.e., both critical points are minima. The result follows by noting that
  $p_1(1; \rho) = p_1(-1;\rho)/9$.
%
 \end{pf}

The following auxiliary results are used in the proof of Theorem~\ref{theo.MSE}.
Recall that $\wh{\delta}^{(h)}$ is the ordinary difference-based estimator
of gap $h$, cf.~\eqref{eq.odb.h}. 
Define
\begin{equation}~\label{eq.CandD}
  {\bf C}_h^{1/2}
  =
  \sum_{i=1}^{n-h}(\varepsilon_i-\varepsilon_{i+h})^2,\quad
  {\bf D}_h^{1/2}
  =
  \sum_{i=0}^{K-1}\sum_{j\in I_{t_i}^{h,0}}\,(a_{i}-a_{i+1})(\varepsilon_j - \varepsilon_{j+h}).
\end{equation}
 We can show that
\begin{equation}\label{eq.Exp-gammahat2-h1}
  (2(n-h))^2\,\ME[ ( \wh{\delta}^{(h)} )^2 ]
  =
  h^2 J_K^2 + 4 h J_K (n-h) (\gamma_0 - \gamma_h) + \ME[{\bf C}_h] + \ME[{\bf D}_h].
\end{equation}
 For $\ME[{\bf C}_h]$ observe that due to Eqs.~\eqref{eq.ExpValE2E2}-\eqref{eq.ExpValE2EE}-\eqref{eq.ExpValEEEE},
 \[
  \ME[ (\varepsilon_i-\varepsilon_{i+h})^2\,(\varepsilon_j-\varepsilon_{j+h})^2 ]
  =
  4(\gamma_0-\gamma_h)^2 + \vt_1(i,j) + \vt_2(i,j)
 \]
 with 
 \[
 \vt_1(i,j)=\mbox{const.}\,\gamma_{|j-i+s|}^2,\quad
 \vt_2(i,j)=\sum_{s,t}\,\mbox{const.}\,\gamma_{|j-i+s|}\gamma_{|j-i+t|},
 \]
 where $s,t\in\{0,\pm h\}$. That is,
\begin{equation}\label{eq.Exp-Ah}
 \ME[ {\bf C}_h ] = [2(n-h)]^{2}(\gamma_0-\gamma_h)^2 + S_{1,n}^{(h)},
\end{equation}
 where $S_{1,n}^{(h)}=\sum_{i,j}^{n-2h}[\vt_1(i,j) + \vt_2(i,j)]$;
 note that $S_{1,n}^{(h)}=\mc O(n)$.

 \begin{lem}\label{lem.exp-val.B}
  Suppose that the conditions of Theorem~\ref{theo.MSE} hold.
  Let ${\bf D}_h$ be defined by \eqref{eq.CandD}. Then, $\ME[{\bf D}_h]=F_h(\gamma_{(\cdot)})\,J_K$ where
  $F_1 = 2(\gamma_0-\gamma_1)$ and for $2\leq h\leq m$
  \begin{equation}\label{eq.exp-val.B}
   F_h(\gamma_{(\cdot)})
   =
   2\left[(h-1)(\gamma_0-\gamma_h) + 
   \sum_{j=2}^h\,\sum_{i=1}^{j+1}\left( 2\gamma_{| j-i |} - \gamma_{| j-i-h |} -\gamma_{|j-i+h|}\right) \right]
  \end{equation}  
 \end{lem}
 \begin{pf}
  For $h=1$ the result follows by noting that for any $t_i$, $I_{t_i}^{h,0}=\{\,t_i-1\,\}$.
  For $2\leq h\leq m$ note that
  \begin{align*}
   {\bf D}_h
   =
   \sum_{i=0}^{K-1}\sum_{k\in I_{t_i}^{h,0}} (a_i-a_{i+1})^2 (\varepsilon_k - \varepsilon_{k+h})^2
   +
   \sum_{s=1}^{K-2}\sum_{r=1}^{K-1}(a_s-a_{s+1})(a_r-a_{r+1})(\tilde{{\bf D}}_{r,s} + \tilde{{\bf D}}_{s,r}),
  \end{align*}
  where 
  \[
  \tilde{{\bf D}}_{s,t}
  =
  \sum_{i\in I_{t_r}^{h,0} }\sum_{j\in I_{t_s}^{h,0} }(\varepsilon_i-\varepsilon_{i+h})(\varepsilon_j-\varepsilon_{j+h}).
  \]
  
  Since for any $t_i$,
  $\ME\left[ \sum_{j\in I_{t_i}^{h,0}}(\varepsilon_j - \varepsilon_{j+h}) \right]^2
  =
  2(h-1)(\gamma_0-\gamma_h)+\Lambda^\ast(h;\gamma_{(\cdot)})$, where
  \[
  \Lambda^\ast(h;\gamma_{(\cdot)})
  = 
  2\,\sum_{j=2}^h\,\sum_{i=1}^{j+1}\left( 2\gamma_{| j-i |} - \gamma_{| j-i-h |} -\gamma_{|j-i+h|}\right),
  \]
  the result is established if we show that $\ME[\tilde{{\bf D}}_{r,s}]=\ME[\tilde{{\bf D}}_{s,r}]=0$.
  To this end, observe that for any $s\in\{1,\ldots,K-2\}$
  and $t\in\{s+1,\ldots,K-1\}$:
  \begin{align*}
   \ME[\tilde{{\bf D}}_{s,r}]
   &=
   \sum_{i=t_s-h}^{t_s-1}\sum_{i=t_r-h}^{t_r-1}\left[ 2\gamma_{|i-j|}-\gamma_{|j-i-h|}-\gamma_{|j-i+h|} \right]\\
   \intertext{let $x=t_r-t_s$ and recall that by assumption, $\min_{1\leq i\leq K-1}|t_i-t_{i-1}|>4(m+1)$ to get,}
   &=
   \sum_{i=1}^h\,\sum_{j=1}^h\,\left[ 2\gamma_{|x +j-i|}-\gamma_{|x+j-i-h|}-\gamma_{|x+j-i+h|} \right]=0.
  \end{align*}
  The last equality follows because $\gamma_{|h|}=0$ for all $h\geq m+1$.
  A similar argument shows that $\ME[\tilde{{\bf D}}_{r,s}]=0$. This completes the proof.
 \end{pf}
 
 From now on, $\sum_{i,j}:=\sum_{i=1}^{n_m}\sum_{j=1}^{n_h}$. 
 \begin{lem}\label{lem.exp-val.prod}
  Supppose that the conditions of Theorem~\ref{theo.MSE} hold.
  Let $\wh{\delta}^{(h)}$ and $\wh{\gamma}_0^{(m)}(d)$ be given by 
  Eqs.~\eqref{eq.odb.h}-\eqref{eq.gamma0}. Then,
  \begin{equation}\label{eq.exp-val.prod}
    \ME[ \wh{\gamma}_0^{(m)}(d)\times \wh{\delta}^{(h)} ]
    =
    \frac{1}{2P(d)n_hn_m}
    \left\{
    [ I^\ast + II^\ast + III^\ast ]\,J_K + S_{2,n}^{(h)}(d)
    \right\},
  \end{equation}  
  where
  \begin{align*}
  I^\ast
  &=
  (m+1)(d^2+1)\,[ 2n_h(\gamma_0-\gamma_h) + h\,J_K]\\
  II^\ast
  &=
  8(d^2-1)V_h,\qquad
  V_h 
  =
  \sum_{s=0}^{m}\sum_{t=1}^h\,\gamma_{s+t} - \sum_{s=1}^{m+1}\sum_{t=1}^h\,\gamma_{|t-s|},
  \quad
  1\leq h\leq m,\\
  III^\ast
  &=
  2P(d)\,h\,n_m\gamma_0. 
  \end{align*}
  Here, $S_{2,n}^{(h)}(d)=\mc O(n)$ and does not depend on $J_K$.
 \end{lem} 
 \begin{pf}   
  By definition
  \[
   \ME[ \wh{\gamma}_0^{(m)}(d) \times \wh{\delta}^{(h)} ]
   =
   \frac{\ME[ I + II + III ]}{2P(d)\,n_h\,n_m},
  \]
  where 
  \begin{align*}
   I
   &=
   \sum_{i,j}\left[ \delta_i^2(d)(\chi_j + \varepsilon_{j}-\varepsilon_{j+h})^2 \right],\quad
   II
   =
   2\sum_{i,j}\,\delta_i(d)\,\eta_i(d)
   \left[ 2\chi_j(\varepsilon_j -\varepsilon_{j+h}) + (\varepsilon_j -\varepsilon_{j+h})^2 \right]\\
   III
   &=
   \sum_{i,j}\eta_i^2(d)\left[ \chi_j + \varepsilon_j -\varepsilon_{j+h} \right]^2
   \end{align*}
   
   Since for all $j$, $\ME[\varepsilon_j]=0$ and
   $\ME[ ( \varepsilon_j-\varepsilon_{j+h} )^2 ]=2(\gamma_0-\gamma_h)$,
   we utilize the arguments leading to Eqs.~\eqref{eq.ExpValGamma0}-\eqref{eq.Exp-gammahat-h1}
   and get
   \begin{equation}\label{eq.exp-val.I}
    I^\ast = \ME[I] 
    =
    (m+1)(d^2+1)\,[ 2n_h(\gamma_0-\gamma_h) + h\,J_K]\,J_K.
   \end{equation}
   
   Due to Gaussianity for any $i$ and $j$, $\ME[ \eta_i(d) (\varepsilon_j - \varepsilon_{j+h})^2 ]=0$.
   Thus,
   according to Lemma~\ref{lem.exp-val.Psi}
   \begin{equation}\label{eq.exp-val.II}
    I^{\ast\ast} = \ME[II]
    =
    4\sum_{i,j}\ME[ \delta_i(d)\,\eta_i(d)\,\chi_j(\varepsilon_j -\varepsilon_{j+h}) ]
    =
    8(d^2-1)\,J_K\,V_h.
   \end{equation}
   
   Gaussianity and Lemma~\ref{lem.ExpValEtai2} yield, $\sum_{i,j} \ME [\eta_i^2(d)\chi_j(\varepsilon_j - \varepsilon_{j+h})] = 0$,
   and $\ME [\eta_i^2(d)]=2P(d)\gamma_0$, respectively. Consequently,
    \begin{equation}\label{eq.exp-val.III}
     I^{\ast\ast\ast} = \ME[III]
     =
     2P(d)\gamma_0\,n_m\,h\,J_K + S_{2,n}^{(h)}(d),
    \end{equation}
  where by stationarity, 
  $S_{2,n}^{(h)}(d) = \sum_{i,j}\ME[ \eta_i^2(d) (\varepsilon_j-\varepsilon_{j+h})^2]=\mc O (n)$.
   In order to get Eq.~\eqref{eq.exp-val.prod} sum up 
   Eqs.~\eqref{eq.exp-val.I}-\eqref{eq.exp-val.II}-\eqref{eq.exp-val.III} and arrange terms. 
 \end{pf}
 
\begin{lem}\label{lem.exp-val.Psi}
 Suppose that the conditions of Theorem~\ref{theo.MSE} hold.
  Let
 \[
  \Psi_{K,d}
  =
  \sum_{i,j}\,\delta_i(d)\,\eta_i(d)\chi_j(\varepsilon_j - \varepsilon_{j+h}).
 \]
 Then,
\begin{equation}\label{eq.exp-val.Psi}
 \ME[\Psi_{K,d}]
 =
 2(d^2-1)J_K\,V_h.
\end{equation}
 See Lemma~\ref{lem.exp-val.prod} for a definition of $V_h$.
 \end{lem}
 \vspace{-.5cm}
\begin{pf}
 Set $c_m = 2(m+1)$.
 Since for given $j$,
 \begin{align*}
    \sum_{i\in I_{t_j}^{c_m,c_m/2}} \delta_i(d)\,\eta_i(d) 
    &=
    d\,(a_j - a_{j+1})\times \sum_{ i\in I_{t_j}^{c_m,c_m/2} }\eta_i:= {\bf E}_{\tau_j}(d),\\    
    \sum_{i\in I_{t_j}^{c_m/2,0}} \delta_i(d)\,\eta_i(d)     
    &=
    (a_j - a_{j+1})\times \sum_{ i\in I_{t_j}^{c_m/2,0} }\eta_i(d) := {\bf F}_{\tau_j}(d),
 \end{align*}
  it follows that $\sum_i\,\delta_i(d)\eta_i(d) = \sum_j\,\left( {\bf E}_{\tau_j}(d) + {\bf F}_{\tau_j}(d) \right)$.\smallskip
  
  Let $\sum_{j,I_{\tau_j}^{L,R}}:=\sum_{j=0}^{K-1}\sum_{i\in I_{\tau_j}^{L,R}}$.
  Note that $\chi_i = (a_j-a_{j+1})\ind_{I_{\tau_j}^{h,0}}(i)$
  and this, in turn, implies that
  $\sum_{i}\chi_i(\varepsilon_i - \varepsilon_{i+h}) 
  = 
  \sum_{j,i}(a_j-a_{j+1})(\varepsilon_i - \varepsilon_{i+h}) := {\bf G}_h$.
  Consequently,  
   \[\Psi_{K,d}
   =
   \sum_{j=0}^{K-1}\left( {\bf E}_{\tau_j}(d) + {\bf F}_{\tau_j}(d)  \right)\times {\bf G}_h
   =
   T_1+T_2+T_3 + U_1+U_2+U_3,
   \]
   where
%
 \begin{align*}
  T_1
	&=
	d\,\sum_{j,I_{t_j}^{(c_m,m/2)}}\,(a_{j}-a_{j+1})\varepsilon_i\,{\bf G}_h,\quad
	T_2
	=
	-d(1+d)\,\sum_{j,I_{t_j}^{(c_m,m/2)}}\,(a_j-a_{j+1})\varepsilon_{i+c_m/2}\,{\bf G}_h\\
	T_3
	&=
	d^2\,\sum_{j,I_{t_j}^{(c_m,m/2)}}\,(a_j-a_{j+1})\varepsilon_{i+c_m}\,{\bf G}_h,\quad
  U_1
	=
	\sum_{j,I_{t_j}^{(m/2,0)}}\,(a_{j}-a_{j+1})\varepsilon_i\,{\bf G}_h,\\
	U_2
	&=
	-(1+d)\,\sum_{j,I_{t_j}^{(m/2,0)}}\,(a_{j}-a_{j+1})\varepsilon_{i+c_m/2}\,{\bf G}_h,\quad
	U_3
	=
	d\,\sum_{j,I_{t_j}^{(m/2,0)}}\,(a_{j}-a_{j+1})\varepsilon_{i+c_m}\,{\bf G}_h.
 \end{align*}
 
 The result follows by computing the expected value of $T$'s and $U$'s.
 Now we compute $\ME[T_1]$ and $\ME[U_1]$, the remaining terms of
 $\ME[\Psi_{K,d}]$ can be treated similarly.  
 In what follows,
 \[
    \sum_{r,s}\,\fancyS_{i,j}^{L_1,R_1,L_2,R_2}    
    :=
    \sum_{r\in I_{t_i}^{L_1,R_1}}\,\sum_{s\in I_{t_j}^{L_2,R_2}}\,\varepsilon_r(\varepsilon_s+\varepsilon_{s+h}).
 \]

 We begin by writing $\ME[T_1]=d\ME[ T_{1,1} + T_{1,2} ]$, where
 $T_{1,1} = \sum_{j=0}^{K-1}(a_{j}-a_{j+1})^2\,\sum_{r,s}\,\fancyS_{j,j}^{c_m,c_m/2,h,0}$
 and 
 $T_{1,2} = \sum_{i=0}^{K-2}\,\sum_{j=i+1}^{K-1}(a_{i}-a_{i+1})\,(a_{j}-a_{j+1})
  \left(
   \sum_{r,s}\,\fancyS_{i,j}^{c_m,c_m/2,h,0}  + \sum_{r,s}\,\fancyS_{j,i}^{c_m,c_m/2,h,0}
  \right)$. Note now that for any $\tau_j$ and due to $m$-dependency,
 \begin{align}
  \ME\left[ \sum_{r,s} \fancyS_{j,j}^{c_m,c_m/2, h, 0} \right]
  =
  \sum_{r=\tau_j-c_m}^{\tau_j-(m+2)}\sum_{s=\tau_j-h}^{\tau_j-1}[ \gamma_{|r-s|} - \gamma_{|r-(s+h)|}]
  =
  \sum_{s=m+2}^{2(m+1)}\sum_{t=1}^{h}\gamma_{|s-t|}
  =
  \sum_{s=m+1}^{m+h}\sum_{t=1}^h\gamma_{|s-t|},\label{eq.exp-val.T11.gamma-part}
 \end{align} 
 These calculations hold independently of the value of $\tau_j$, and consequently
 we get that $\ME[T_{1,1}]=\sum_{r=m+1}^{m+h}\sum_{s=1}^h\gamma_{|r-s|}\,J_K$.
 Similar calculations along with the $m$-dependency and \eqref{eq.DistBetweenJumps}
 allows us to get that $\ME[T_{1,2}]=0$. All in all, we have shown that
 \begin{equation}\label{eq.exp-val.T1}
  \ME[T_1]=d\,\sum_{s=m+1}^{m+h}\sum_{t=1}^h\,\gamma_{|h-(s+t)|}\,J_K.
 \end{equation}
 
 Similar arguments yield,
 \begin{align}
  \ME[T_2]
  &=
  -d(1+d)\sum_{s=1}^{m+1}\sum_{t=1}^h\,[ \gamma_{|t-s|} - \gamma_{| t-(s+h) |} ]\,J_K \label{eq.exp-val.T2}\\
  \ME[T_3]
  &=
  d^2\,\sum_{s=0}^{m}\sum_{t=1}^h\,[ \gamma_{s+t} - \gamma_{| h-(s+t) |} ]\,J_K.\label{eq.exp-val.T3}
 \end{align}

 Now, we consider $\ME[U_1]$. Write $U_1=U_{1,1}+U_{1,2}$,
 where
 \begin{align*}
  U_{1,1}
  &=
  \sum_{j=0}^{K-1}(a_j-a_{j+1})^2 \sum_{r,s}\,\fancyS_{j,j}^{c_m/2,0,h,0} \\
  U_{1,2}
  &=
  \sum_{i=0}^{K-2}\,\sum_{j=i+1}^{K-1}(a_{i}-a_{i+1})\,(a_{j}-a_{j+1})
  \left(
    \sum_{r,s}\fancyS_{i,j}^{c_m/2,0,h,0} + \sum_{r,s}\fancyS_{j,i}^{h,0,c_m/2,0}
  \right)
 \end{align*}
 Following \eqref{eq.exp-val.T11.gamma-part} we can show 
 that for any $\tau_j$,
 $\ME\left[ \sum_{r,s} \fancyS_{j,j}^{c_m/2,0,h,0} \right]
  =
  \sum_{s=1}^{m+1}\sum_{t=1}^{h}[ \gamma_{|t-s|} -\gamma_{| h-(t+s) |} ]$.
 Again $m$-dependency and \eqref{eq.DistBetweenJumps} allow us to show that
 $\ME[U_{1,2}]=0$. Therefore,
 \begin{equation}\label{eq.exp-val.U1}
  \ME[U_1]
  =
  \sum_{r=1}^{m+1}\sum_{s=1}^{h}[ \gamma_{|r-s|} -\gamma_{| s-(r+h) |} ]\,J_K.
 \end{equation}

   Similar arguments yield,
 \begin{align}
  \ME[U_2]
  &=
  -(1+d)\sum_{r=0}^{m+1}\sum_{s=1}^h\,[ \gamma_{r+s} - \gamma_{| h-(r+s) |} ]\,J_K \label{eq.exp-val.U2}\\
  \ME[U_3]
  &=
  -d\,\sum_{r=m+1}^{m+h}\sum_{s=1}^h\,\gamma_{| h-(r+s) |}\,J_K.\label{eq.exp-val.U3}
 \end{align}

 Since $\sum_{r=0}^m\sum_{s=1}^h\,\gamma_{|h-(r+s)|}
  =
  \sum_{r=1}^{m+1}\sum_{s=1}^h\,\gamma_{|s-r|}$, 
  $\sum_{r=1}^{m+1}\sum_{s=1}^h\,\gamma_{|s-(r+h)|}
  =
  \sum_{r=0}^{m}\sum_{s=1}^h\,\gamma_{r+s}$, 
  \eqref{eq.exp-val.Psi} follows after summing up \eqref{eq.exp-val.T1}-\eqref{eq.exp-val.U3}.
\end{pf}

\section{Proofs and auxiliary results for Section~\ref{sec:HolderRegression}}~\label{sec.AppendixIII}

In this Appendix we will assume that the conditions of Theorem~\ref{theo.AsympProperties.Gammah}
hold and use the notation introduced at the beginning of Appendices~\ref{sec.AppendixII} and 
\ref{sec.Proofs.bias.minimizer}.
Also throughout this section $c_m = 2(m+1)$ and the symbols $\kappa_1$, $\kappa_2$, etc., 
will denote constants which do not depend on $n$. We will write $\sum_{i,j}$ to denote
$\sum_{i=1}^{n_m-1}\,\sum_{i=j+1}^{n_m}$.

\begin{lem}\label{lem.BIAS.Holder}
  Suppose that the conditions of Theorem~\ref{theo.AsympProperties.Gammah} hold.
  Then, for $h = 0,\ldots,m$,
 \[
  \ME[ \wh{\gamma}_h^{(m)}(d_{h,m}) ]
  =
  \gamma_h + \mc O( S_n ),
  \qquad
  S_n
  =
  \sum_{j=1}^{K_n}\,\frac{\vt_j^2}{n}
  +
  \sum_{j=1}^{K_n}\,n^{-2(\alpha_j + 1/2)}.
 \]
\end{lem}
\begin{pf}
 We begin with the case $h=0$.
 Since $\wh{\gamma}_0^{(m)}(d) = (2(d^2+d+1)\,n_m)^{-1}\sum_{i=1}^{n_m}(\delta_i(d) + \eta_i(d))^2$,
 see Appendix~\ref{sec.Proofs.bias.minimizer} for definition of $\delta_i(d)$ and $\eta_i(d)$,
 and $\ME[ \eta_i^2(d) ] = 2(d^2 + d + 1)\gamma_0$, cf.~Corollary~\ref{cor.ExpEtai2},
 in order to analyze the asymptotic bias of $\wh{\gamma}_0^{(m)}(d)$ it suffices to
 focus on $\sum_i\,\delta_i^2(d)$. To this end we write
 $\delta_i^2(d) = s_0^2(i) + 2d\,s_0(i)\,s_{c_m}(i) + d^2\,s_{c_m}^2(i)$ and note that
 for given $i$ there exists a unique $\tau_j$ such that
 \begin{equation*}
  s_0(i)
  =
  \begin{cases}
   t(i,j) := a_j(i/n) - a_j( (i+m+1)/n ) & \mbox{ for } i\in I_{\tau_{j-1}, \tau_j}^{(0, c_m/2)}\\ 
   u(i,j) := a_j(i/n) - a_{j+1}( (i+m+1)/n ) & \mbox{ for } i\in I_{\tau_j}^{(c_m/2, 0)}\\
  \end{cases},
 \end{equation*}
  and a similar characterization holds for $s_{c_m}(\cdot)$, see Eq.~\eqref{eq.Intervals}
  in Appendix~\ref{sec.Proofs.bias.minimizer} for definition of 
  $I_{\tau_{j-1}, \tau_j}^{(0, c_m/2)}$ and $I_{\tau_j}^{(c_m/2, 0)}$.
  This implies that the dominat terms in $\delta_i^2(d)$ are of the form $t^2(i,j) + u^2(i,j)$
  and in what follows we provide bounds for these terms.

  Recall that 
  $\vt_j = a_j(\tau_{j+1}^{-}) - a_{j+1}(\tau_{j+1})$ and by assumption
  there exists a number $c>0$ such that $|\vt_j| > c $ for all $j$. 
  Utilizing the H\"older condition of $f$ we get
  \begin{equation}~\label{eq.boundT}
    \sum_{j=1}^{K_n-1}\,\sum_{i\in I_{\tau_{j-1}, \tau_j}^{(0, c_m/2)}}\,t^2(i,j)
    \leq 
    \kappa_{1,m}\,\sum_{j=1}^{K_n-1}\,n^{-2\alpha_j},   
  \end{equation}
   as well as
  \begin{align*}
  \sum_{j=1}^{K_n-1}\,\sum_{i\in I_{\tau_j}^{(c_m/2, 0)}}\,u^2(i,j)
    &\leq
    \sum_{j=1}^{K_n}\vt_j^2
    +
    \kappa_{2,m}\sum_{j=1}^{K_n}\, n^{-\alpha_j}\,\vt_j
    +
    \kappa_{1,m}\,\sum_{j=1}^{K_n}\,n^{-2\alpha_j}.
  \end{align*}
  Here $\kappa_{1,m}=\sup_j\,(m+1)^{\alpha_j}<\infty$, $\kappa_{2,m}=2\,\kappa_{1,m}$.
  Define the set of indices
  $I_{K_n}:=\{ j\in \{1,\ldots, K_n\}: |\vt_j| \geq 1 \}$ and observe that
  \begin{equation}~\label{eq.boundU}
    \big| \sum_{j=1}^{K_n} n^{-\alpha_j}\,\vartheta_j \big|
    \leq
    \{\sum_{j\in I_{K_n}}\vt_j^2 + \sum_{j\notin I_{K_n}}\,|\vt_j|\}
    \leq 
    \{\sum_{j\in I_{K_n}}\vt_j^2 + \sum_{j\notin I_{K_n}}\frac{1}{c}\,\vt_j^2\}
    \leq
    (1+c^{-1})\,\sum_{j=1}^{K_n}\,\vt_j^2.
  \end{equation}
  From \eqref{eq.boundT} and \eqref{eq.boundU} follows that
  $\sum_{i=1}^{n_m}\delta_i^2(d) = \mc O \left( \sum_{j=1}^{K_n}\,{\vt_j^2} 
  + \sum_{j=1}^{K_n}\,n^{-2\alpha_j}\right)$.
  Consequently,
 \begin{equation}\label{eq.exp-val.gamma0-Lipschitz}
  \ME[ \wh{\gamma}_{0}^{(m)}(d) ]
  =
  \gamma_0 + \mc O \left( S_n \right),
  \quad 
  S_n=\sum_{j=1}^{K_n}\frac{\vt_j^2}{n} + \sum_{j=1}^{K_n}\,n^{-2(\alpha_j + 1/2)}.
 \end{equation}
 
 For $h \geq 1$, firstly note that by writing 
 $\wh{\delta}^{(h)} = (2(n-h))^{-1}\sum_{i=1}^{n_h}(s_0(i) + \eta_i(0))^2$
 we can mimick the calculations above and get
 \begin{equation}
 \ME [ \wh{\delta}^{(h)} ]
 =
 \gamma_0 - \gamma_h
 +
 \mc O(S_n).~\label{eq.ExpVal-Rh}  
 \end{equation}
 Then, by definition, $\wh{\gamma}_h^{(m)}(d) = \wh{\gamma}_0^{(m)}(d) - \wh{\delta}^{(h)}$, cf.~\eqref{eq.gammak}
 in the Introduction, and the result follows by adding Eqs.~\eqref{eq.exp-val.gamma0-Lipschitz} and \eqref{eq.ExpVal-Rh}.  
\end{pf}

\begin{lem}\label{lem.VAR.Holder}
 Suppose that the assumptions of Lemma~\ref{lem.BIAS.Holder} hold.
 Then
 \begin{equation}\label{eq.VAR.gamma0}
  \VAR( \wh{\gamma}_0^{(m)}(d) )
  =
  \mc O ( \sum_{j=1}^{K_n} (\vt_j^2/n^2 + n^{-(\alpha_j + 2)} ) )
  +
  \mc O (n^{-1})
 \end{equation}
 the same result holds for $\VAR( \wh{\delta}^{(h)} )$. Moreover,
 \begin{equation}\label{eq.VAR.gammah}
    \VAR( \wh{\gamma}_h^{(m)}(d) ) = \mc O ( \VAR( \wh{\gamma}_0^{(m)}(d) ) +    
     ( \sum_{j=1}^{K_n}\,{|\vt_j|}/{n} )^2 + ( \sum_{j=1}^{K_n}\,n^{-(\alpha_j+1)} )^2 ).  
 \end{equation}
\end{lem}
\begin{pf}
 We write $\wh{\gamma}_0^{(m)}(d)=n^{-1}\sum\,b_i^2(d)$, where $b_i(d) = \delta_i(d) + \eta_i(d)$,
 see Appendix~\ref{sec.Proofs.bias.minimizer} for notation. It is easily seen that
 $\VAR (b_i^2(d)) = 4\delta_i^2(d)\,\VAR(\eta_i(d)) + \VAR(\eta_i^2(d))$. 
 From Corollary~\ref{cor.ExpEtai2} and Lemma~\ref{lem.ExpValEtai4}, $\VAR(\eta_i(d))$
 and $\VAR(\eta_i^2(d))$ are uniformly bounded. This implies that the order 
 of magnitude of $\sum_i\,\VAR(b_i^2(d))$ depends solely on $\sum_i\,\delta_i^2(d)$. From arguments in the proof
 of Lemma~\ref{lem.BIAS.Holder} we get that
 \begin{equation}~\label{eq.Var.bi2}
  \sum_{i=1}^{n_m}\,\VAR( b_i^2(d) )
  =
  \mc O \left( \sum_{j=1}^{K_n}\vt_j^2 + \sum_{j=1}^{K_n}\,n^{-2\alpha_j}  \right)
  + \mc O(n). 
 \end{equation}
 
 It can be shown that 
 $\COV(b_i^2(d), b_j^2(d))
  =
  4\delta_i(d)\delta_j(d)\ME[ \eta_i(d)\,\eta_j(d) ]
  +
  2\delta_i(d)\ME[ \eta_i(d)\,\eta_j^2(d) ]
  +
  2\delta_j(d)\ME[ \eta_j(d)\,\eta_i^2(d) ]
  +
  \COV( \eta_i^2(d), \eta_j^2(d) )$.
  Due to $m$-dependency and stationarity of moments up to 4-th order
  we get for any $i$ and $j$, $\ME[ \eta_i(d)\,\eta_j(d) ] = \kappa_2\,\mu_2(|j-i|)$,
  $\ME[ \eta_i(d)\,\eta_j^2(d) ] = \kappa_3\,\mu_3(|j-i|)$ and
  $\COV( \eta_i^2(d), \eta_j^2(d) ) = \kappa_4\,\mu_4(|j-i|)$, where
  $\mu_2(\cdot)$, $\mu_3(\cdot)$ and $\mu_4(\cdot)$ are functions which depend only 
  on sums of moments of second, third and fourth order of $\varepsilon$, respectively.
  With the same arguments used in Lemma~\ref{lem.A}, we can establish
  that $\sum_{i,j}\,\delta_i(d)\,\delta_j(d)\mu_2(j-i) = \mc O (n)$.
  Similar arguments allow us to get that $\sum_{i,j}\delta_i(d)\,\mu_3(j-i) = \mc O ( \sum_i\,\delta_i(d) )$
  and $\sum_{i,j}\,\COV( \eta_i^2(d), \eta_j^2(d) ) = \mc O (n)$.
  
  All in all, we have proven that
  \begin{equation}~\label{eq.Cov.bi2.bj2}
    \sum_{i,j}\,\COV(b_i^2(d), b_j^2(d))
    =
    \mc O\left( \sum_{i}\,\delta_i(d) \right) + \mc O (n).
  \end{equation}
  Since $\delta_i(d) = s_0(i) + ds_{c_m}(i)$, see Appendix~\ref{sec.Proofs.bias.minimizer}
  for notation, from the characterization of $s_0(\cdot)$ and $s_{c_m}(\cdot)$ given in Lemma~\ref{lem.BIAS.Holder}
  it follows that 
  the order of magnitude of \eqref{eq.Cov.bi2.bj2} is driven
  by 
  $\sum_{i}\,s_0(i)$. 
  We get,
  \begin{equation}
   \bigg| \sum_{i=1}^{n_m}s_0(i) \bigg|
   =
   \bigg| \sum_{i=1}^{K_n-1}\left(
   \sum_{j\in I_{\tau_{i-1}, \tau_i}^{(0, c_m/2)}}\,t(j,i)
   +
   \sum_{j\in I_{\tau_i}^{(c_m/2, 0)}}\,u(j,i)   
   \right) \bigg|
   =
   \mc O \left( \sum_{j=1}^{K_n} (n^{-\alpha_j} + |\vt_j|^2) \right).\label{eq.VAR.Aux2}
  \end{equation}
  The latter follows because of H\"older condition on $f$
  and calculations leading to \eqref{eq.boundU}.
  Observe that Eq.~\eqref{eq.VAR.gamma0} follows by a combination of 
  Eqs.~\eqref{eq.Var.bi2}-\eqref{eq.Cov.bi2.bj2}-\eqref{eq.VAR.Aux2}.
 
 For $h\geq 1$, we write $\wh{\delta}^{(h)}= (2(n-h))^{-1}\sum_{i=1}^{n_h}(s_0(i) + \eta_i(0))^2$
 and mimick the calculations above to deduce that $\wh{\gamma}_0^{(m)}(d)$
 and $\wh{\delta}^{(h)}$ have variances of the same order.
 Then, since $\wh{\gamma}_h^{(m)} = \wh{\gamma}_0^{(m)} - \wh{\delta}^{(h)}$, cf.~\eqref{eq.gammak}
 in the Introduction, we combine Eq.~\eqref{eq.VAR.gamma0} and Lemma~\ref{lem.COV.Holder} below to show
 the validity of Eq.~\eqref{eq.VAR.gammah}. This completes the proof.
\end{pf}

\begin{lem}\label{lem.COV.Holder}
 Suppose that the assumptions of Lemma~\ref{lem.BIAS.Holder} hold.
 Then
 \begin{equation}\label{eq.CovGeneral}
  \COV( \wh{\gamma}_0^{(m)}(d_{h,m}), \delta^{(h)} )
  =
  \mc O \left(  ( \sum_{j=1}^{K_n}\,{|\vt_j|}/{n} )^2 + ( \sum_{j=1}^{K_n}\,n^{-(\alpha_j+1)} )^2 \right) + \mc O (n^{-1}).      
 \end{equation}
\end{lem}
\begin{pf}
  We begin with the case $d_{h,m}=1$.
  Throughout this proof $k \in \{h, c_m\}$.
  By definition, 
 \begin{equation}\label{eq.lem.COV.HOlder.Aux1}
  \COV( \wh{\gamma}_0^{(m)}(1), \wh{\delta}^{(h)} )
  =
  \frac{1}{12\,n_m\,n_h}\,
  \sum_{i=1}^{n_m}\,\sum_{j=1}^{n_h}\,\COV( z_{i, c_m},z_{j,h} ) + \mc O(n^{-1}),  
 \end{equation}
 where for given index $i$, $z_{i,k} = y_{i:(i+k+1)}^\top\,D_{k+2}\,y_{i:(i+k+1)}$,
 see Appendix~\ref{sec.AppendixII} for notation of $y_{i:(i+k+1)}$
 and Eq.~\eqref{eq.Dtilde} for definition of the $(k+2)\times (k+2)$ matrix $D_{k+2}$.
 We also write $c_i(k) = f_{i:(i+k+1)}^\top\,D_{k+2}\,\varepsilon_{i:(i+k+1)}$
 and $d_{i}(k) = \varepsilon_{i:(i+k+1)}^\top\,D_{k+2}\,\varepsilon_{i:(i+k+1)}$
 which by standard calculations yield
 $\COV( z_{i, c_m}, z_{j, h} ) =  4\ME [ c_i(c_m)\,c_j(h) ] + 
 2\ME[ c_i(c_m)\,d_j(h) ] + 2\ME[ c_j(h)\,d_i(c_m) ] - 
 \mbox{tr}( D_{c_m+2}\,\Sigma_{c_m+2} )\,\mbox{tr}( D_{h+2}\,\Sigma_{h+2} )$.
  Stationarity and $m$-dependency, arguments also used in Lemma~\ref{lem.VAR.Holder}, 
  allow us to get that the second, third and fourth summands above are sums of stationary
  moments of second, third and fourth order, respectively. Hence the contribution
  of these terms to \eqref{eq.lem.COV.HOlder.Aux1} is of order $\mc O (n^{-1})$.
  It is not difficult to see that for given indeces $i$ and $j$, 
  $c_i(m)\,c_j(h)$  is the sum of 8 terms of the form
  $(f_i-f_{i+(m+1)})(f_j-f_{j+h})( \varepsilon_i - 2\varepsilon_{i+m+1} 
  + \varepsilon_{i+2(m+1)} )(\varepsilon_j - \varepsilon_{j+h})$
  and due to stationarity and $m$-dependency, $\ME [ c_i(m)\,c_j(h) ]$
  is bounded by 
 $|f_i-f_{i+(m+1)})(f_j-f_{j+h})|$. Now, 
 since $s_k(i) = f_{i+k} - f_{i+m+1}$, see notation in Appendix~\ref{sec.Proofs.bias.minimizer},
 we utilize the ideas leading to the bound of $\VAR ( \wh{\gamma}_0^{(m)}(1)  )$, cf.~\eqref{eq.VAR.gamma0},
 and obtain that
 \begin{equation}\label{eq.lem.COV.HOlder.Aux3}
 \big| \sum_{i=1}^{n_m}\,\sum_{j=1}^{n_h}\,(f_i-f_{i+(m+1)})(f_j-f_{j+h}) \big|
 =
 \mc O \left( ( \sum_{j=1}^{K_n}|\vt_j| )^2 + ( \sum_{j=1}^{K_n}\,n^{-\alpha_j} )^2  \right).
 \end{equation}
 
 Thus for $d_{h,m}=1$, the result follows by a combination 
 of Eqs.~\eqref{eq.lem.COV.HOlder.Aux1} and \eqref{eq.lem.COV.HOlder.Aux3}.
 For the other values of $d_{h,m}$, cf.~\eqref{eq.Delta(121).2} in the Introduction, we  
 mimick the calculations above to complete the proof.
\end{pf}

\bibliographystyle{apalike}
\bibliography{MultiscaleBib}

%
%
%
%
%
\end{document}